\documentclass[letterpaper,english,prl,aps,10pt]{revtex4-1}
\usepackage[T1]{fontenc}
\usepackage{amsmath}
\usepackage{amssymb}

\makeatletter

\pdfpageheight\paperheight
\pdfpagewidth\paperwidth

\usepackage{dcolumn}
\usepackage{bm}

\usepackage{graphicx}			
\usepackage{enumerate}
\usepackage[shortlabels]{enumitem}
\usepackage{amssymb}
\usepackage{subfigure}
\usepackage{amsthm}
\usepackage{multirow}
\usepackage{booktabs}
\usepackage{amsmath,scalerel}
\usepackage{rotating}
\usepackage[utf8]{inputenc}

\DeclareMathOperator*{\Bigcdot}{\scalerel*{\cdot}{\bigodot}}
\usepackage[colorlinks=true]{hyperref}
\hypersetup
{
colorlinks,%
citecolor=green,%
linkcolor=blue,%
urlcolor=blue,%
}
\usepackage{tikz}

\usepackage{braket}
\renewcommand{\v}[1]{\ensuremath{\mathbf{#1}}} 
\newcommand{\gv}[1]{\ensuremath{\text{\boldmath$ #1 $}}}
\newcommand{\abs}[1]{\left| #1 \right|} 
\newcommand{\trace}[1]{\mathrm{tr}#1} 
\let\baraccent=\= 
\renewcommand{\=}[1]{\stackrel{#1}{=}} 

\newcommand{\thmref}[1]{\hyperref[#1]{Theorem~\ref{#1}}}
\newcommand{\figref}[1]{\hyperref[#1]{Fig.~\ref{#1}}}
\renewcommand{\eqref}[1]{\hyperref[#1]{Eq.~(\ref{#1})}}
\newcommand{\appref}[1]{\hyperref[#1]{Appendix~(\ref{#1})}}

\makeatother

\usepackage{babel}
\begin{document}

\title{A modern description of Rayleigh's criterion}

\author{Sisi Zhou}

\affiliation{Departments of Applied Physics and Physics, Yale University, New Haven,
Connecticut 06511, USA}
\affiliation{Yale Quantum Institute, Yale University, New Haven, Connecticut 06520, USA}

\author{Liang Jiang}

\affiliation{Departments of Applied Physics and Physics, Yale University, New Haven,
Connecticut 06511, USA}
\affiliation{Yale Quantum Institute, Yale University, New Haven, Connecticut 06520, USA}

\date{\today}

\begin{abstract}
Rayleigh's criterion states that it becomes essentially difficult to resolve two incoherent optical point sources separated by a distance below the width of point spread functions (PSF), namely in the subdiffraction limit. Recently, researchers have achieved superresolution for two incoherent point sources with equal strengths using a new type of measurement technique, surpassing Rayleigh's criterion. However, situations where more than two point sources needed to be resolved have not been fully investigated. Here we prove that for any incoherent sources with arbitrary strengths, a one- or two-dimensional image can be precisely resolved up to its second moment in the subdiffraction limit, i.e. the Fisher information (FI) is non-zero. But the FI with respect to higher order moments always tends to zero polynomially as the size of the image decreases, for any type of non-adaptive measurement. We call this phenomenon a modern description of Rayleigh's criterion. 
For PSFs under certain constraints, the optimal measurement basis estimating all moments in the subdiffraction limit for 1D weak-source imaging is constructed. Such basis also generates the optimal-scaling FI with respect to the size of the image for 2D or strong-source imaging, which achieves an overall quadratic improvement compared to direct imaging. 
\end{abstract}
\maketitle

\section{\label{sec:intro}Introduction}

Rayleigh's criterion, as a long-standing textbook theorem, puts a fundamental limit on the power of optical resolution \cite{rayleigh1879xxxi,born2013principles}. It states that when two points are separated from each other by a distance smaller than the width of point-spread function (PSF) of the optical system, namely in the subdiffraction limit, it becomes essentially difficult to distinguish them. Recently however, researchers made a breakthrough towards surpassing Rayleigh's criterion using a new type of measurement technique, by looking at the imaging problem from the perspective of quantum metrology \cite{tsang2016quantum,ang2017quantum,nair2016far,lupo2016ultimate,chrostowski2017superresolution,rehacek2017multiparameter,lu2018quantum,yu2018quantum}. 

In metrology, Fisher information (FI) characterizes the ultimate precision of parameter estimation through Cram\'er-Rao bound \cite{kobayashi2011probability,helstrom1976quantum,braunstein1994statistical}. When estimating the separation between two equal strength incoherent sources, it can be shown that FI tends to zero as they become closer when we use direct imaging approach (i.e. counting photons at different positions on the imaging plane). However, the quantum Fisher information (QFI, equal to the maximum FI over all possible quantum measurements) remains a constant, implying the possibility of superresolution \cite{tsang2016quantum}. In fact, many types of measurement have been proposed to achieve this kind of superresolution \cite{tsang2016quantum,nair2016interferometric,ang2017quantum,nair2016far,lupo2016ultimate,yang2017fisher,rehacek2016dispelling,tsang2017subdiffraction,parniak2018superresolution} and some of these approaches have already been demonstrated experimentally \cite{tang2016fault,yang2016far,tham2017beating,paur2016achieving}. For example, when the PSF is Gaussian, it is possible to achieve the highest estimation precision by projecting the optical field onto Hermite-Gaussian modes \cite{tsang2016quantum,tsang2017subdiffraction,rehacek2016dispelling}.  

While this new approach appears to be a promising candidate to substantially improve imaging resolution, many questions are yet to be answered: (1) What is the ultimate precision one can achieve, in a general imaging scenario, given experimentalists access to all types of measurement? (2) Which type of measurement achieves such precision? In this paper, we tackle these two questions by conducting a comprehensive Fisher information analysis 
in the general scenario where the incoherent source distribution on the object plane 
is arbitrary. 

A direct way to parametrize an image is to use positions and intensities of each point as parameters to be estimated. However, it may not be the perfect choice because the position of one specific point does not tell much about the structure of the whole image. Instead, we can use moments to characterize an image which has wide applications in image analysis \cite{flusser20162d}. Since the difficulty involved in calculating QFI increases significantly as the number of sources increase, we only consider the limiting values of QFIs as the size of the image tends to zero (much smaller than the width of PSF) which we call ``the subdiffraction limit''. 

In this paper, we choose normalized moments (normalized so that it has dimension of length) as parameters to be estimated, where detailed calculations for Gaussian PSFs and the spatial-mode demultiplexing (SPADE) measurement scheme are contained in Refs.~\cite{tsang2017subdiffraction,chrostowski2017superresolution}. We obtain the fundamental precision limit of estimating moments in the subdiffraction limit which formulated a modern description of Rayleigh's criterion, as opposed to the traditional Rayleigh's criterion restricted by direct imaging. We find that the FI with respect to (wrt) second moment remains a positive value in the subdiffraction limit, in accordance with previous work on estimating the separation between two coherent source. However, the FI wrt higher order moments always vanishes in the subdiffraction limit for non-adaptive measurements, answering question (1). This result shows the capability of going beyond direct imaging will not provide unlimited power and only push image resolution one step forward -- from the first moment (the centroid of the image) to the second moment. To be specific, if we use $s$ to represent the size of an image, the FI wrt to the $K$-th order moment vanishes as $O(s^{K-2})$ ($O(s^{K-1})$) when $K$ is even (odd), compared with $O(s^{2K-2})$ using direct imaging.    

Based on the FI analysis, we also obtain optimal quantum measurements (in the subdiffraction limit) corresponding to the optimal FI. It is shown in this paper that when PSF is under certain constraints, the optimal measurement basis is strongly related to its derivatives. Roughly speaking, the probability from projecting the optical field onto the $K$-th order derivative of the PSF provides information of the $2K$-th order moment of the image. And choosing derivatives as the measurement basis successfully classifies information of different moments into different measurement outcomes, which will provide optimal FIs wrt these moments in the subdiffraction limit. In this paper, we partially answer question (2) by first providing optimal quantum measurement scheme for second moment. 
For higher order moments, we prove the optimality of this scheme for 1D weak-source imaging. For 2D imaging or for strong-source imaging, such scheme only provides the optimal scaling of FI wrt $s$, but the coefficient may be further improved.

\newpage
\section{\label{sec:summary}Summary of results}

Here we briefly summarize our results on Fisher information analysis for incoherent optical imaging. 
\begin{itemize}
    \item In \autoref{sec:formalism}, we provide the formalism of the far-field imaging of incoherent optical sources, where we use P representation of optical states to express the Fisher information matrix (FIM). 
    \item In \autoref{sec:weak}, we consider imaging for weak incoherent sources in one-dimensional imaging. We show that the Fisher information (FI) with respect to normalized moments decreases polynomially as the size of the image decrease, by order-of-magnitude analysis. To be specific, the FI wrt second moments remains a constant as the size of the image tends to zero, and the FI wrt to higher order moments drops to zero. 
    \item In \autoref{sec:strong}, we generalize the statement in \autoref{sec:weak} to sources with arbitrary strength, again by order-of-magnitude analysis.
    \item In \autoref{sec:second}, we detail the FI analysis wrt to second moments by providing the exact value of FI and corresponding optimal measurements, as FI wrt second moment is not influenced by Rayleigh's criterion.
    \item In \autoref{sec:2D}, we generalize all discussions about one-dimensional imaging to two-dimensional imaging, including calculating FI wrt to second moments in 2D. 
    \item In \autoref{sec:all-moments}, we detail the FI analysis wrt to all moments and show how the optimal scaling of FI can be achieved wrt all moments, which is improved \emph{quadratically} when compared to direct imaging. \autoref{sec:all-moments} also serves as a justification of the order-of-magnitude analysis in \autoref{sec:weak} and \autoref{sec:strong}. 
\end{itemize}

We also summarize the contents of each appendix here: 
\begin{itemize}
    \item \appref{app:convergence} discusses the condition under which the series expansion of probabilities and FIs. For a well-behaved point spread function, the series expansion of probability converges uniformly and therefore the FIs can also be expanded wrt different orders of the size of the image. We also point out that our analysis can only be applied to non-adaptive measurements in order for the series expansion to be valid. 
    \item \appref{app:expansion} provides the first three terms in the series expansion of measurement probability for arbitrary incoherent sources, which is not explicitly given in \autoref{sec:strong}. 
    \item \appref{app:angle} provides an alternative way to parametrize second moments in 2D imaging, as opposed to the one in \autoref{sec:2D}. 
    \item \appref{app:odd-moments}, \appref{app:2D} and \appref{app:generalization} complement discussions in \autoref{sec:all-moments} in terms of optimizing FI wrt odd moments for weak incoherent sources in 1D imaging, 2D imaging, generalization to arbitrary strengths.
    \item \appref{app:centroid} discusses the pre-estimation of the centroid. We provide a measurement scheme which is optimal for weak sources and at least 96.4\% efficient for strong sources. 
\end{itemize}


\newpage 
The main results in this paper are also summaried in \autoref{tb:sum-1D} and \autoref{tb:sum-2D} for further reference.
\begin{table}[ht]
\footnotesize
\begin{center}
\begin{tabular}{ |c|c|c| } 
\hline
& Weak source ($\epsilon \ll 1$) & Strong source (arbitrary $\epsilon$) \\
\hline
\multirow{2}{*}{Moments} & \multicolumn{2}{c|}{$M_{k} = \big(\sum_{j} \gamma_j (x_j - \bar X)^k\big)^{1/k}$} \\
 & \multicolumn{2}{c|}{\eqref{eq:moment-1D}} \\
\hline
\multirow{5}{*}{Probability for outcome $n$} & \multicolumn{2}{c|}{$P(n;\{x_j,\Gamma_j\}) = \mathbb E[\braket{\psi_\alpha|E(n)|\psi_\alpha}]$} \\
 & \multicolumn{2}{c|}{\eqref{eq:prob}} \\
\cline{2-3}
& $P(n;\{x_j,\Gamma_j\}) = (1-\epsilon)\braket{0|E(n)|0} + \epsilon p(n) + O(\epsilon^2)$ & $P(n;\{x_j,\Gamma_j\}) = \sum_{k=0}^\infty Q_k(n;\{M_\ell,\ell\leq k\})$ \\
& \eqref{eq:weak-qfi} & \eqref{eq:strong-qfi} \\
\cline{2-3}
& $p(n) = \sum_{k=0}^\infty \frac{p_k(n)}{k!}(M_k)^k$ & \multirow{2}{*}{\appref{app:expansion}}\\
& \eqref{eq:weak-qfi-series} & \\
\hline
\multirow{1}{*}{FI} & \multicolumn{2}{c|}{$\mathcal{F}_{k\ell}=\sum_{n}\frac{1}{P(n;\{x_{j},\Gamma_{j}\})}\Big(\frac{\partial P(n;\{x_{j},\Gamma_{j}\})}{\partial M_{k}}\Big)^2=O(s^{k-2})\qquad$\eqref{eq:fi}} \\
\hline
\multirow{3}{*}{Maximum FI} &  \multicolumn{2}{c|}{$\max_{\{E(n)\}}\mathcal F_{kk} = \begin{cases} O(s^{k-2}) \quad k \text{ is even,}\\ O(s^{k-1})\text{ is odd.}\end{cases}$}\\
\cline{2-3}
 & $\max_{\{E(n)\}}\lim_{s\rightarrow 0}\mathcal F_{2\ell\,2\ell} = \epsilon q_{\ell}^2(2\ell)^2 (M_{2\ell})^{2\ell-2} $ & \multirow{2}{*}{$\max_{\{E(n)\}}\lim_{s\rightarrow 0}\mathcal F_{22} = 4\epsilon \Delta k^2$} \\
 & \eqref{eq:1D-even} & \\
 & $\max_{\{E(n)\}}\lim_{s\rightarrow 0}\mathcal F_{2\ell+1\,2\ell+1} = 4\epsilon q_{\ell+1}^2(2\ell+1)^2 \frac{(M_{2\ell+1})^{4\ell}}{(M_{2\ell})^{2\ell}} $& \multirow{2}{*}{\eqref{eq:fi2}}\\
 & \eqref{eq:1D-odd} & \\
\hline
\multirow{2}{*}{Optimal Measurement} & $B_0^{\rm w}$, $B_1^{\rm w}$ and $B_2^{\rm w}$ & For $M_2$, $E(N_{0})=\sum_{k=0}^\infty \frac{(\psi_{{\bar X}}^{\dagger})^k\psi_{{\bar X}}^{(1)\dagger}\ket{0}\bra{0}\psi_{{\bar X}}^{(1)}(\psi_{{\bar X}})^{k}}{k!\braket{0|\psi_{{\bar X}}^{(1)}\psi_{{\bar X}}^{(1)\dagger}|0}}   $\\
& \autoref{sec:all-moments} &  \eqref{eq:strong-measurement} \& \appref{app:generalization}\\
\hline
\end{tabular}
\caption {\label{tb:sum-1D} A summary of the main results (1D)}
\end{center}
\end{table}

\begin{table}[ht]
\footnotesize
\begin{center}
\begin{tabular}{ |c|c|c| } 
\hline
& Weak source ($\epsilon \ll 1$) & Strong source (arbitrary $\epsilon$) \\
\hline
\multirow{2}{*}{Moments} & \multicolumn{2}{c|}{$M_{k\ell} = \big(\sum_{j} \gamma_j (x_j - \bar X)^k(y_j - \bar Y)^\ell\big)^{1/(k+\ell)}$} \\
 & \multicolumn{2}{c|}{\eqref{eq:moment-2D}} \\
\hline
\multirow{5}{*}{Probability for outcome $n$} & \multicolumn{2}{c|}{$P(n;\{x_j,y_j,\Gamma_j\}) = \mathbb E[\braket{\psi_\alpha|E(n)|\psi_\alpha}]$} \\
 & \multicolumn{2}{c|}{\eqref{eq:prob}} \\
\cline{2-3}
& $P(n;\{x_j,y_j,\Gamma_j\}) = (1-\epsilon)\braket{0|E(n)|0} + \epsilon p(n) + O(\epsilon^2)$ & \multirow{4}{*}{ $P(n;\{x_j,y_j,\Gamma_j\}) = \sum_{K=0}^\infty Q_K(n;\{M_{k\ell},\ell + k\leq K\})$} \\
& \eqref{eq:weak-qfi} & \\
\cline{2-2}
& $p(n) = \sum_{k\ell=0}^\infty \frac{p_{k\ell}(n)}{k!\ell!}(M_{k\ell})^{k+\ell}$ & \\
& $p_{k\ell}(n) = \partial_{\bar X}^k\partial_{\bar Y}^\ell \braket{0|\psi_{\bar X\bar Y}E(n)\psi_{\bar X\bar Y}^\dagger|0}$ & \\
\hline
\multirow{1}{*}{FI} & \multicolumn{2}{c|}{$
\mathcal{F}_{k\ell\,k'\ell'}=\sum_{n}\frac{1}{P(n;\{x_{j},y_{j},\Gamma_{j}\})}\Big(\frac{\partial P(n;\{x_{j},y_{j},\Gamma_{j}\})}{\partial M_{k\ell}}\Big)^2=O(s^{k+\ell-2})\qquad,$\eqref{eq:fi-2}} \\
\hline
\multirow{2}{*}{Maximum FI} &  \multicolumn{2}{c|}{$\max_{\{E(n)\}}\mathcal F_{L\,(K-L)\;L\,(K-L)} = \begin{cases} O(s^{K-2}) \quad k \text{ is even,}\\ O(s^{K-1})\text{ is odd.}\end{cases}$}\\
 & \multicolumn{2}{c|}{\autoref{sec:all-moments} \& \appref{app:2D}}\\
\hline
\multirow{2}{*}{Optimal Measurement} & $B_{0,1,2,3,4,5,6}^{\rm w}$ & For $M_{20},M_{11}$ and $M_{02}$, see \autoref{sec:2D}.\\
& \autoref{tb:2D} & \appref{app:generalization}\\
\hline
\end{tabular}
\caption {\label{tb:sum-2D} A summary of the main results (2D)}
\end{center}
\end{table}
\newpage

\section{\label{sec:formalism}Formalism}

Consider a one-dimensional object composed of $J$ points on the object plane. The original field on the object plane can be expressed using P representation \cite{walls2007quantum},
\begin{equation}
\rho_{0}=\int D\alpha P_{\Gamma_0}(\alpha)\ket\alpha\bra\alpha,
\end{equation}
where $\alpha=(\alpha_{1},\ldots,\alpha_{J})^{T}$ is the column vector
of complex field amplitudes for $J$ optical spatial modes and 
\begin{equation}
\ket\alpha=\Big(\prod_{j=1}^{J}e^{-\frac{|\alpha_{j}|^{2}}{2}}e^{\alpha_{j}a_{j}^{\dagger}}\Big)\ket0,
\end{equation}
where $\ket0$ is the vacuum state, $a_{j}^{\dagger}$ and $a_{j}$
are the canonical creation and annihilation operators at position
$x_{j}$. Suppose the fields are uncorrelated at different points
on the object plane, then $P_{\Gamma_0}(\alpha)$
is the independent Gaussian distribution of the $J$ modes:
\begin{equation}
P_{\Gamma_0}(\alpha)=\prod_{j=1}^{J}\frac{1}{\pi(\Gamma_0)_{j}}\exp\Big(-\sum_{j=1}^{J}\frac{|\alpha_{j}|^{2}}{(\Gamma_0)_{j}}\Big),
\end{equation}
where $(\Gamma_0)_{j}\geq 0$ is the average photon number emitted at the
$j$th point and $\Gamma_0=((\Gamma_0)_{1},\ldots,(\Gamma_0)_{J})^{T}$. 

The imaging system maps the source operators $a_{j},a_{j}^{\dagger}$
into the image operators $\psi_{j},\psi_{j}^{\dagger}$ with an attenuation
factor $\eta$: 
\begin{equation}
a_{j}^{\dagger}\rightarrow\sqrt{\eta}\psi_{j}^{\dagger}+\sqrt{1-\eta}v_{j}^{\dagger}.
\end{equation}
Here $\eta$ is the transmission probability. $\psi_{j}^{\dagger}=\int dx \psi_{\mathrm{PSF}}(x-x_{j})a_{x}^{\dagger}$
is described by the point-spread function $\psi_{\rm PSF}(x)$ (normalized) where $a_{x}^{\dagger}$
is the canonical creation operator at position $x$ and $v_{j}^{\dagger}$
is the creation operator of the auxiliary environmental modes 
\cite{lupo2016ultimate}. 
Moreover, we assume the PSF satisfies the following assumption 
\begin{equation}
\label{eq:prop-1D}
\int_{-\infty}^\infty \Big(\frac{d^{\ell}}{dx^{\ell}}\psi^*_{\mathrm{PSF}}(x)\Big)\Big(\frac{d^{\ell+1}}{dx^{\ell+1}}\psi_{\mathrm{PSF}}(x)\Big) dx= 0, \;\forall \ell \geq 0.
\end{equation}
which will later be used in determining the optimal measurement basis. This assumption is easily satisfied, for example, when PSFs are real (real PSFs can be implemented by a two-lens system \cite{goodman2005introduction}), e.g. $\psi_{\mathrm{PSF}}(x) \propto e^{-x^2/4\sigma^2}$; or when they are even, e.g. $\psi_{\mathrm{PSF}}(x) \propto e^{ikx^2/2z}{\rm sinc}(x/\sigma)$.

The field on the image plane is expressed as 
\begin{equation}
\begin{split}
\rho &= \trace_{\rm env}\Big(\int D\alpha P_{\Gamma_0}(\alpha)\Big(\prod_{j=1}^{J}e^{-\frac{|\alpha_{j}|^{2}}{2}}e^{\alpha_{j}\psi_{j}^{\dagger}}e^{\sqrt{1-\eta}\alpha_{j}v_{j}^{\dagger}}\Big)\ket0 \bra0 \Big(\prod_{j=1}^{J}e^{-\frac{|\alpha_{j}|^{2}}{2}}e^{\alpha_{j}\psi_{j}}e^{\sqrt{1-\eta}\alpha_{j}v_{j}} \Big)\Big)\\
&= \int D\alpha P_{\Gamma}(\alpha) \ket{\psi_\alpha}\bra{\psi_\alpha} ,
\end{split}
\end{equation}
where 
\begin{equation}
\ket{\psi_{\alpha}} = \frac{\prod_{j=1}^{J}e^{-\frac{|\alpha_{j}|^{2}}{2}}e^{\alpha_{j}\psi_{j}^{\dagger}}\ket0}{\big(\bra{0}\prod_{j=1}^{J}e^{-|\alpha_{j}|^{2}}e^{\alpha_{j}^*\psi_{j}} e^{\alpha_{j}\psi_{j}^{\dagger}}\ket{0}\big)^{1/2}},
\end{equation} 
and $\Gamma := \eta \Gamma_0$ is the average photon number received from each mode. We also define the average photon number on the image plane $\epsilon := \sum_{j=1}^J \Gamma_{j} $ (which is usually a small number) and the relative source strength $\gamma_j := \Gamma_j/\epsilon$ for later use. We can see that after integrating all phases in $\alpha$, only those photon number diagonal terms will survive and we may write
\begin{equation}
\rho = \sum_{{\rm m}=0}^\infty \pi_{\rm m} \rho_{\rm m}
\end{equation}
where $\pi_{\rm m}$ is the probability of having $\rm m$ photons in the
state and $\rho_{\rm m}$ is an $\rm m$-photon multimode Fock state.

Our goal is to extract information of the image from $\rho$. We use
a set of positive operators $\{E(n)\}$ satisfying $\sum_{n}E(n)=I$
to represent the positive-operator valued measure (POVM) performed on $\rho$ \cite{nielsen2010quantum,helstrom1976quantum}.
The resultant probability distributions are 
\begin{equation}
P(n;\{x_{j},\Gamma_{j}\})=\trace({\rho E(n)})\equiv\mathbb{E}[\braket{\psi_{\alpha}|E(n)|\psi_{\alpha}}],
\end{equation}
where $\mathbb{E}[\cdot]$ represents expectation values under Gaussian distribution $P_{\Gamma}(\alpha)$.

The Cram\'er-Rao bound \cite{kobayashi2011probability}
\begin{equation}
\label{eq:CRbound}
\Sigma\succeq\mathcal{F}^{-1}
\end{equation}
provides the ultimate precision limit in terms of parameter estimation, where ``$\succeq$'' means the LHS minus the RHS is positive semi-definite, $\Sigma_{k\ell}$ is
the error covariance matrix wrt parameters $\{M_{k}\}_{k\geq1}$
and
\begin{equation}
\mathcal{F}_{k\ell}=\sum_{n}\frac{1}{P(n;\{x_{j},\Gamma_{j}\})}\frac{\partial P(n;\{x_{j},\Gamma_{j}\})}{\partial M_{k}}\frac{\partial P(n;\{x_{j},\Gamma_{j}\})}{\partial M_{\ell}}
\end{equation}
is the corresponding Fisher information matrix (FIM). $M_k$ are some functions of $\{x_j,\Gamma_j\}$, later chosen to be the normalized moments.

\section{\label{sec:weak}The ultimate resolution limit for weak incoherent sources}

The probability of measurement outcome $n$ is
\begin{equation}
\label{eq:prob}
P(n;\{x_{j},\Gamma_{j}\})=\mathbb{E}[\braket{\psi_{\alpha}|E(n)|\psi_{\alpha}}]=\mathbb{E}\Big[\frac{\braket{0|e^{\alpha^{\dagger}\psi}E(n)e^{\psi^{\dagger}\alpha}|0}}{\braket{0|e^{\alpha^{\dagger}\psi}e^{\psi^{\dagger}\alpha}|0}}\Big],
\end{equation}
where $\psi=(\psi_{1},\ldots,\psi_{J})^{T}$ is the column vector
of annihilation operators $\psi_{j}$. In the limit where the average
photon number on the image plane $\epsilon$ is small (the value of $\epsilon$ is considered known because it is easy to measure), we can expand
it as a series in $\epsilon$:
\begin{equation}
P(n;\{x_{j},\Gamma_{j}\})=(1-\epsilon)\braket{0|E(n)|0}+\epsilon \,p(n) +O(\epsilon^{2}),
\label{eq:weak-qfi}
\end{equation}
where
$p(n) := \epsilon \sum_{j=1}^{J}\gamma_{j}\braket{0|\psi_{j}E(n)\psi_{j}^{\dagger}|0}.$
Since the first term contains no information of the object, the FIM
will be dominated by the second term, which corresponds to the situation
where only one photon is detected. To study the behavior of FIM in the subdiffraction limit, we expand $\psi_{j}$ around its centroid $\bar X$. 
One should be careful with the convergence radius of the series expansion though, which has a lower bound independent of the measurement $E(n)$ (see \appref{app:convergence}). 
The second term in \eqref{eq:weak-qfi} becomes
\begin{equation}
\epsilon \,p(n)
=\epsilon\sum_{k=0}^{\infty}\frac{p_k(n)}{k!}(M_{k})^{k},
\label{eq:weak-qfi-series}
\end{equation}
where $p_k(n) = \sum_{j=1}^{J} \big(\frac{\partial}{\partial x_j}\big)^k p(n)\big|_{x_j = \bar X}$ is equal to the $k$-th order derivative of $\braket{0|\psi_{{\bar X}}E(n)\psi_{{\bar X}}^{\dagger}|0}$ wrt ${\bar X}$ and $M_{k}$ are normalized moments defined by 
\begin{equation}
\label{eq:moment-1D}
M_{k}=\Big(\sum_{j=1}^{J}\gamma_{j}(x_{j}-{\bar X})^{k}\Big)^{1/k}
\end{equation}
for $k\geq 0$. Note that $(\Bigcdot)^{1/k}$ is introduced here only to make sure $M_k$ has dimension of length so that the estimation error should be comparable with the size of the image.  
Other methods to normalize moments, e.g. $M_k = (\sum_{j=1}^{J}\gamma_{j}(x_{j}-{\bar X})^{k})/(\sum_{j=1}^{J}\gamma_{j}(x_{j}-{\bar X})^{2})^{\frac{(k-1)}{2}}$ should also generate similar results. Here we wouldn't worry about the phase of $M_{k}$ because it is well defined locally. For example when $M_k = i\abs{M_k}$, we can estimate $\abs{M_k}$ instead so that all parameters are real. 

Although $\{M_{k}\}_{k\geq1}$ fully characterize the object configuration, they may not be independent given prior information of the object, but we can always choose a smaller set of independent moments as the parameters to be measured. For example, if the object contains only two points, there are only three degrees of freedom --- the positions of two points and the ratio of their strengths, then we choose the first three moments as the parameters to be measured. 

We use $s=\max_{i,j} \abs{x_j - x_i}$ to characterize the size of the image and conduct FI analysis in the subdiffraction limit when $s\rightarrow 0$. 
Here we assume the centroid of the image $\bar X = \sum_{j=1}^J \gamma_j x_j$ is known accurately either based on existing telescopic data or pre-estimation. 
In this case, we have $M_1 = 0$. In \appref{app:centroid}, we provide a measurement scheme for pre-estimation of $\bar X$. In 1D imaging, the scheme is optimal for weak sources and at least 96.4\% efficient for strong sources. The methodology behind this scheme is not clear until \autoref{sec:all-moments}. Therefore we are not going to explain it in detail here.  


Since any converging power series is dominated by its first non-zero
term as $s\rightarrow0$, we have
\begin{equation}
\frac{\partial P(n;\{x_{j},\Gamma_{j}\})}{\partial M_{k}}=O(s^{k-1})\;\;\text{and}\;\;\frac{1}{P(n;\{x_{j},\Gamma_{j}\})}\frac{\partial P(n;\{x_{j},\Gamma_{j}\})}{\partial M_{k}}=O(s^{-1}).\label{eq:order}
\end{equation}
Note that when the terms of lower order than $k$ in $P(n;\{x_{j},\Gamma_{j}\}$ does not vanish, $\frac{1}{P(n;\{x_{j},\Gamma_{j}\})}\frac{\partial P(n;\{x_{j},\Gamma_{j}\})}{\partial M_{k}}$ should be bounded by a power of $s$ with higher order than $O(s^{-1})$. From \eqref{eq:order}, the FI for $k \geq 2$ would be
\begin{equation}
\mathcal{F}_{kk}=\sum_{n}\frac{1}{P(n;\{x_{j},\Gamma_{j}\})}\bigg(\frac{\partial P(n;\{x_{j},\Gamma_{j}\})}{\partial M_{k}}\bigg)^2=O(s^{k-2}),\label{eq:fi} 
\end{equation}
which indicates the following theorem:

\vspace{0.1in}
\noindent\textbf{Theorem 1 (Modern Rayleigh's criterion for one-dimensional imaging)}: \emph{
For imaging of incoherent point sources in the subdiffraction limit, the estimation variance of moment $M_{k>2}$ increases inverse-polynomially as $s$ decreases;
meanwhile, the estimation variance of the second moment $M_{2}$ is bounded by a constant independent of $s$.}
\vspace{0.1in}

Note that we only need to bound the diagonal element of the FIM because the variance in estimation $M_k$ satisfies
\begin{equation}
\Sigma_{kk} \geq (\mathcal F^{-1})_{kk} \geq \mathcal F_{kk}^{-1}.
\end{equation}
where the equality holds true when $\mathcal F$ is diagonal. 

\begin{figure}[htbp] 
\subfigure[\label{fig:1D-dist}]{\includegraphics[width=8.5cm]{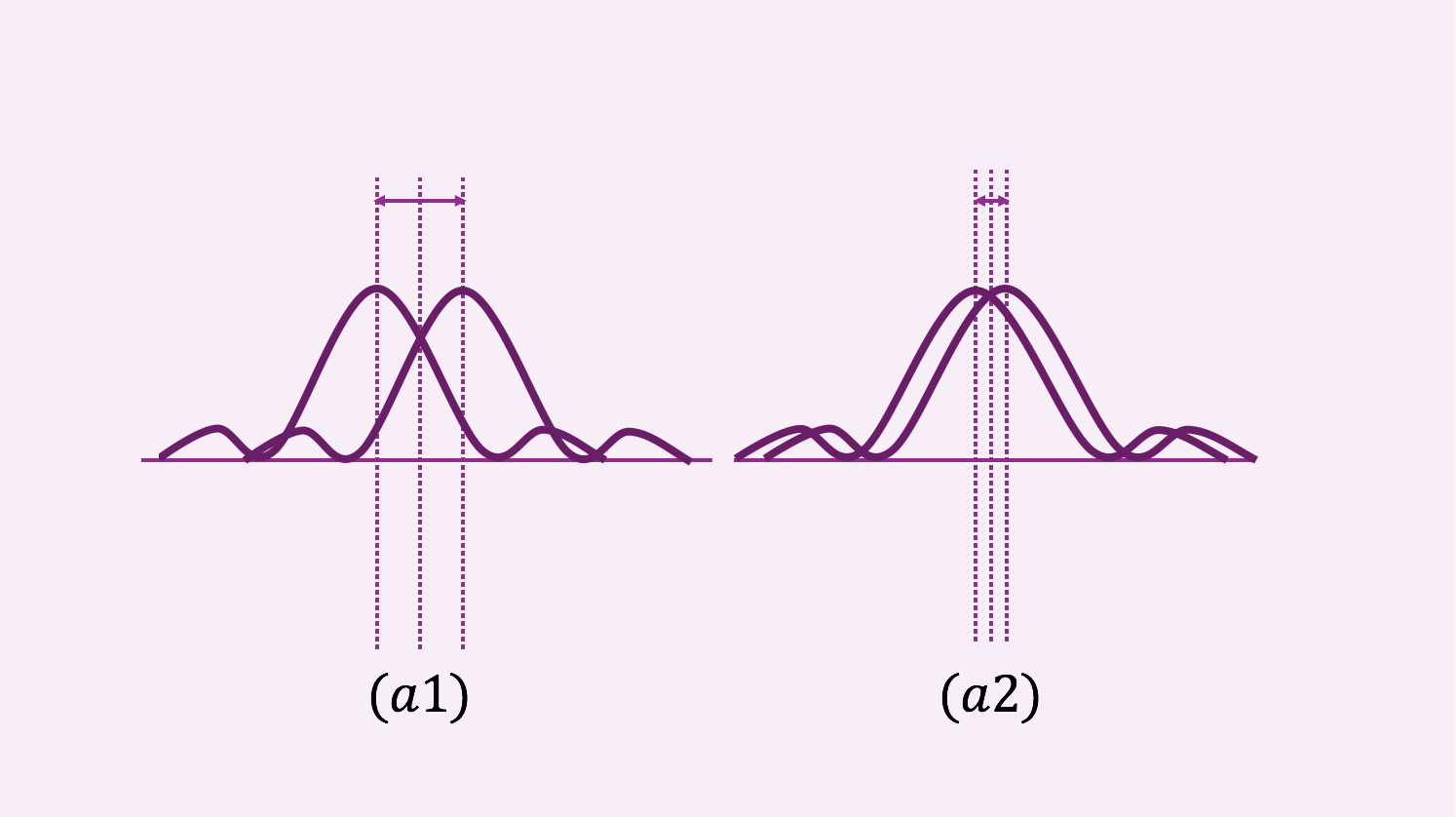}
}
\hspace{0.2cm}
\subfigure[\label{fig:1D-indist}]{\includegraphics[width=8.5cm]{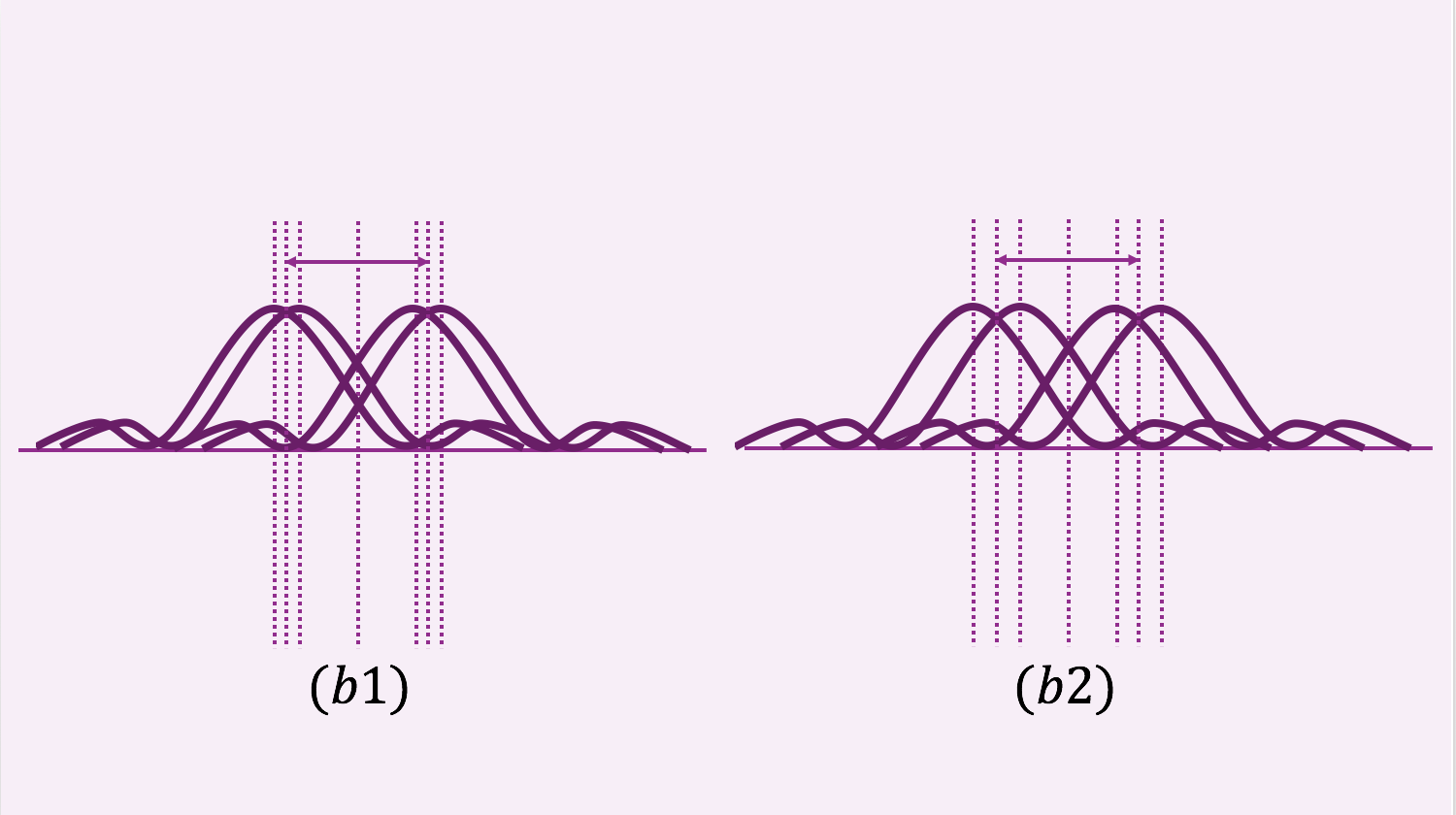}
}
\caption{\label{fig:1D} (a)~Images (a1) and (a2) have different $M_2$. Consider two point sources with equal source strengths. The distance between them equal to $2M_2$ can be measured precisely, therefore it shall be easy to distinguish (a1) and (a2). (b)~Images (b1) and (b2) have the same $M_2$ but different $M_4$. Consider four point sources with equal source strengths. It is difficult to estimate the third and higher moments to distinguish the two images from each other.}
\end{figure}

A simple schematic illustration of above theorem is shown in \autoref{fig:1D}. Further justifications are contained in \autoref{sec:strong}, \autoref{sec:second} and \autoref{sec:all-moments}. 
We discuss the validity of this order-of-magnitude analysis in \appref{app:convergence}. 
{We emphasize here that the measurement is assumed to be non-adaptive in this paper and 
our analysis does not include the case where measurement can be adaptively modified (\appref{app:convergence}) assuming prior knowledge on the moments to be estimated. 
And the adaptive measurement is excluded because it requires demanding experimental techniques. A more general analysis through direct calculation of quantum Fisher information, which can be applied to all type of measurement, can be found in Ref.~\cite{tsang2018quantum}.

}

\section{\label{sec:strong}The ultimate resolution limit for incoherent sources with arbitrary
strengths}

In this section, we generalize the above discussion in weak source limit 
to sources with arbitrary strengths. 

In \eqref{eq:prob}, we replace $\psi^{\dagger}\alpha$ with its
expansion $\sum_{j=1}^J \alpha_{j} \int dx \psi_{\mathrm{PSF}}(x-x_j)a_x^\dagger \equiv \sum_{k=0}^{\infty}\frac{A^{(k)}}{k!}\psi_{{\bar X}}^{(k)\dagger}$,
where $A^{(k)}=\sum_{j=1}^{J}\alpha_{j}(x_{j}-{\bar X})^{k}$ and 
\begin{equation}
\psi_{{\bar X}}^{(k)\dagger}=\frac{d^{k}}{d{\bar X}^{k}}\int dx \psi_{\mathrm{PSF}}(x-{\bar X})a_{x}^{\dagger}.
\end{equation}
According to Wick's theorem (Isserlis' theorem) \cite{isserlis1918formula}, any moment of Gaussian distributions can be calculated using the values of second order moments 
\begin{equation}
\label{eq:gaussian-second-moment}
\mathbb{E}[A^{(\ell_1)}A^{(\ell_2)*}] = \sum_{j=1}^J \Gamma_{j} (x_j-{\bar X})^{\ell_1+\ell_2}; \qquad \mathbb{E}[A^{(\ell_1)}A^{(\ell_2)}] = 0.
\end{equation}
Here $\mathbb{E}[A^{(\ell_1)}A^{(\ell_2)}]$ vanishes when integrating wrt phases of $\alpha$.

We observe that $P(n;\{x_{j},\Gamma_{j}\})$ can be decomposed into a power
series of $O(s)$, like in \eqref{eq:weak-qfi-series},
\begin{equation}
P(n;\{x_{j},\Gamma_{j}\})=\sum_{k=0}^{\infty}Q_{k}(n;\{M_{\ell},\ell\leq k\}),
\label{eq:strong-qfi}
\end{equation}
where $Q_{k}(n;\{M_{\ell},\ell\leq k\})$ is a function of the moments $M_{\ell}$ with $\ell \leq k$ 
so that $Q_{k}(n;\{M_{\ell}\})=O(s^{k})$. Explicit expressions of $Q_{0,1,2}(n)$ are provided in \appref{app:expansion}. 
For example, 
\begin{equation}
\label{eq:q0}
Q_{0}(n)=\sum_{k=0}^{\infty}\frac{\epsilon^{k}}{(1+\epsilon)^{k+1}}\braket{0|(\psi_{{\bar X}})^{k}E(n)(\psi_{{\bar X}}^{\dagger})^{k}|0},
\end{equation}
which is the probability of outcome $n$ when all $J$ points are
located at the centroid ${\bar X}$ with thermal average `excitation' number $\epsilon$. Hence, we have shown that order-of-magnitude analysis is still valid. 

Specially, for $\epsilon \ll 1$, the expansion of $Q_k(n)$ depends solely on $p_k(n)$ and $M_k$:
\begin{equation}
Q_0(n) = \bra{0}E(n)\ket{0} + O(\epsilon),\quad
Q_{k}(n) = \epsilon\,\frac{p_k(n)}{k!} (M_k)^k + O(\epsilon^2), \,\forall k\geq 1,
\end{equation}
and \eqref{eq:strong-qfi} simplifies to \eqref{eq:weak-qfi} for weak incoherent sources. 
We also notice that $Q_2(n)/Q_0(n) = O(\epsilon s^2)$ (see \appref{app:expansion}), which means the subdiffraction limit (requiring $Q_2(n)\ll Q_0(n)$) needs smaller $s$ as $\epsilon$ increases.  


\section{\label{sec:second}FI wrt second moment and corresponding optimal measurement}

In \autoref{sec:weak}, we have shown that there is a possibility to obtain a
non-zero FI wrt $M_{2}$. We are now going to find the exact value of the optimal FI wrt second moment and corresponding measurement basis. First, let's consider the weak-source scenario, 
\begin{equation}
\mathcal{F}_{22}=\sum_{n}\frac{1}{P(n;\{x_{j},\Gamma_{j}\})}\bigg(\frac{\partial P(n;\{x_{j},\Gamma_{j}\})}{\partial M_{2}}\bigg)^{2}.
\end{equation}
As $s\rightarrow0$, $P(n;\{x_{j},\Gamma_{j}\})$ and $\frac{\partial P(n;\{x_{j},\Gamma_{j}\})}{\partial M_{2}}$
will be dominated by its first non-zero term, therefore according
to \eqref{eq:weak-qfi-series},
\begin{equation}
\lim_{s\rightarrow0}\mathcal{F}_{22}=\epsilon\sum_{n\in N^{\rm w}_{0}}\frac{1}{\frac{p_2(n)}{2}(M_{2})^{2}}\big(p_2(n)(M_{2})\big)^{2}=4\epsilon\big(\braket{0|\psi_{{\bar X}}^{(1)}E(N^{\rm w}_{0})\psi_{{\bar X}}^{(1)\dagger}|0}+\mathrm{Re}[\braket{0|\psi_{{\bar X}}^{(2)}E(N^{\rm w}_{0})\psi_{{\bar X}}^{\dagger}|0}]\big),
\end{equation}
where {we define a set of $0$-null measurement outcomes} $N^{\rm w}_{0}=\{n|\braket{0|E(n)|0} = \braket{0|\psi_{{\bar X}}E(n)\psi_{{\bar X}}^{\dagger}|0}=0\}$
and $E(N^{\rm w}_{0})=\sum_{n\in N^{\rm w}_{0}}E(n)$. We also note that $p_0(n) = 0$ implies $p_1(n) = 0$. Since $E(N^{\rm w}_0)$ is Hermitian and non-negative, its eigenstates corresponding to non-vanishing eigenvalues must be orthogonal to $\psi_{{\bar X}}^{\dagger}|0\rangle$ and $\mathrm{Re}[\braket{0|\psi_{{\bar X}}^{(2)}E(N^{\rm w}_{0})\psi_{{\bar X}}^{\dagger}|0}]$ must be zero. Therefore,
\begin{equation}
\label{eq:fi2-weak}
\max_{\{E(n)\}}\lim_{s\rightarrow0}\mathcal{F}_{22}=4\epsilon\big(\braket{0|\psi_{{\bar X}}^{(1)}\psi_{{\bar X}}^{(1)\dagger}|0}=4\epsilon\int |\partial_x \psi_{\mathrm{PSF}}(x)|^{2}dx \equiv 4\epsilon \Delta k^2,
\end{equation}
where the first equality is achieved when $\psi_{{\bar X}}^{(1)\dagger}|0\rangle$
is an eigenstate of $E(N^{\rm w}_{0})$ with an eigenvalue equal to one. For example, 
\begin{equation}
\label{eq:weak-measurement}
E(N^{\rm w}_{0})=\frac{\psi_{{\bar X}}^{(1)\dagger}\ket{0}\bra{0}\psi_{{\bar X}}^{(1)}}{\braket{0|\psi_{{\bar X}}^{(1)}\psi_{{\bar X}}^{(1)\dagger}|0}}
\end{equation}
is optimal, in accordance with the optimality of the SPADE measurement scheme for Gaussian PSFs \cite{tsang2016quantum}. Furthermore, if $\psi_{\mathrm{PSF}}(x)$ is an even function, its derivative will be odd and we can also choose $E(N^{\rm w}_0)$ to be $\frac{I-\mathcal{P}}{2}$ where $\mathcal{P}$ is the parity operator satisfying $\mathcal P\cdot f(x) = f(-x)$, which is the so-called SLIVER measurement scheme \cite{nair2016interferometric}. This type of measurement does not depend on the specific expressions of the point-spread functions. 

We emphasize that above discussions are only applicable in the subdiffraction limit and the optimal measurement should be modified for finite $s$. When we consider the special case where there are only two equal strength point sources, however, \eqref{eq:weak-measurement} remains optimal even when $s$ is large \cite{tsang2016quantum}. 

When we use direct imaging approach, i.e. $\{E(n)\}=\{a_{x}^{\dagger}a_{x}dx\}$,
the 0-null measurement outcomes have zero measure and $\lim_{s\rightarrow0}\mathcal{F}_{22}=0$,
because the probability density of the photon position $x$ is
\begin{equation}
\braket{0|\psi_{{\bar X}}a_{x}^{\dagger}a_{x}\psi_{{\bar X}}^{\dagger}|0}=|\psi_{\mathrm{PSF}}(x-{\bar X})|^{2}\neq0\quad\text{almost everywhere},
\end{equation}
which explains the traditional Rayleigh's criterion. 

For an arbitrary source strength
\begin{equation}
\lim_{s\rightarrow0}\mathcal{F}_{22}=\sum_{n\in N_{0}}\frac{1}{Q_{2}(n)}\bigg(\frac{\partial Q_{2}(n)}{\partial M_{2}}\bigg)^{2},
\end{equation}
where the 0-null measurement outcome $N_{0}=\{n|Q_{0}(n)=\sum_{k=0}^{\infty}\frac{\epsilon^{k}}{k!(1+\epsilon)^{k+1}}\braket{0|(\psi_{{\bar X}})^{k}E(n)(\psi_{{\bar X}}^{\dagger})^{k}|0}=0\}=\{n|\braket{0|(\psi_{{\bar X}})^{k}E(n)(\psi_{{\bar X}}^{\dagger})^{k}|0}=0,\;\forall k\}$. We also note that $Q_0(n)=0$ implies $Q_1(n) = 0$ (see \appref{app:expansion}).
A detailed calculation of \eqref{eq:prob} shows that when $n\in N_0$,
\begin{equation}
\label{eq:q2}
Q_{2}(n)=\Big(\sum_{k=0}^{\infty}\frac{\epsilon^{k+1}}{k!(1+\epsilon)^{k+1}}\braket{0|(\psi_{{\bar X}})^{k}\psi_{{\bar X}}^{(1)}E(n)\psi_{{\bar X}}^{(1)\dagger}(\psi_{{\bar X}}^{\dagger})^{k}|0}\Big)M_{2}^{2},
\end{equation}
and hence
\begin{equation}
\label{eq:fi2}
\max_{\{E(n)\}}\lim_{s\rightarrow0}\mathcal{F}_{22}=4\epsilon\int|\partial_x \psi_{\mathrm{PSF}}(x)|^{2}dx = 4\epsilon\Delta k^2.
\end{equation}
It has the exact same expression as \eqref{eq:fi2-weak}, meaning FI wrt the second moment grows linearly as the source strength grows, following the standard quantum limit \cite{giovannetti2004quantum}. Our results agree with previous work on estimating the separation between two incoherent sources for arbitrary source strengths \cite{nair2016far,lupo2016ultimate}. 

The measurement is optimal when $(\psi_{{\bar X}}^{\dagger})^k\psi_{{\bar X}}^{(1)\dagger}|0\rangle$
are all eigenstates of $E(N_{0})$ with eigenvalues equal to one.
For example, 
\begin{equation}
\label{eq:strong-measurement}
E(N_{0})=\sum_{k=0}^\infty \frac{(\psi_{{\bar X}}^{\dagger})^k\psi_{{\bar X}}^{(1)\dagger}\ket{0}\bra{0}\psi_{{\bar X}}^{(1)}(\psi_{{\bar X}})^{k}}{k!\braket{0|\psi_{{\bar X}}^{(1)}\psi_{{\bar X}}^{(1)\dagger}|0}}
\text{  (or when $\psi_{\rm PSF}(x)$ is even, } 
E(N_{0})=\frac{1-\mathcal P}{2})
\end{equation}
is optimal, in accordance with the optimality of fin-SPADE and pix-SLIVER \cite{nair2016far}.

\section{\label{sec:2D}Generalization to two-dimensional imaging}

Results in previous sections can be directly generalized to two-dimensional imaging. Suppose there are $J$ point sources at positions $(x_j,y_j)$. The normalized moments are redefined as following:
\begin{equation}
\label{eq:moment-2D}
M_{k\ell} = \bigg(\sum_{j=1}^J \gamma_j (x_j-{\bar X})^k (y_j- {\bar Y})^\ell\bigg)^{\frac{1}{k+\ell}}
\end{equation}
which fully characterizes the object configuration. Also, the size of the image $s := \max_{ij}\sqrt{(x_i-x_j)^2+(y_i-y_j)^2}$ and the centroid $({\bar X}, {\bar Y}) := (\sum_{j=1}^J \gamma_j x_j, \sum_{j=1}^J \gamma_j y_j)$. 
We can expand the creation and annihilation operators around the centroid ($\partial^{k}_{{\bar X}}$ denotes the $k$-th order derivative wrt $\bar X$)
\begin{equation}
\begin{split}
\psi_j^\dagger &= \int dxdy\psi_{\mathrm{PSF}}(x-x_j,y-y_j) a_{xy}^\dagger \\
&= \sum_{k,\ell=0}^\infty \frac{\int dxdy\partial_{{\bar X}}^k\partial_{ {\bar Y}}^\ell \psi_{\mathrm{PSF}}(x-{\bar X},y- {\bar Y}) a_{xy}^\dagger}{k!\ell!}  (x_j-{\bar X})^k (y_j- {\bar Y})^\ell 
\equiv  \sum_{k,\ell=0}^\infty \frac{\psi^{(k\ell)\dagger}_{{\bar X} {\bar Y}}}{k!\ell!}  (x_j-{\bar X})^k (y_j- {\bar Y})^\ell,
\end{split}
\end{equation}
and calculate the probability distribution $P(n;\{x_{j},y_{j},\Gamma_{j}\})$ which is a series of $O(s^k)$
\begin{equation}
P(n;\{x_{j},y_{j},\Gamma_{j}\})=\sum_{K=0}^{\infty}Q_{K}(n;\{M_{k \ell},k+\ell \leq K\}).
\end{equation}
Similar to \eqref{eq:order}, we have the following order-of-magnitude analysis:
\begin{equation}
\frac{\partial P(n;\{x_{j},y_{j},\Gamma_{j}\})}{\partial M_{k\ell}}=O(s^{k+\ell-1})\;\;\text{and}\;\;\frac{1}{P(n;\{x_{j},y_{j},\Gamma_{j}\})}\frac{\partial P(n;\{x_{j},y_{j},\Gamma_{j}\})}{\partial M_{k\ell}}=O(s^{-1});\label{eq:order-2}
\end{equation}
and similar to \eqref{eq:fi}, the diagonal elements of the FI matrix is 
\begin{equation}
\mathcal{F}_{k\ell\,k\ell}=\sum_{n}\frac{1}{P(n;\{x_{j},y_{j},\Gamma_{j}\})}\bigg(\frac{\partial P(n;\{x_{j},y_{j},\Gamma_{j}\})}{\partial M_{k\ell}}\bigg)^2=O(s^{k+\ell-2}),\label{eq:fi-2}
\end{equation}
thus extending the modern description of Rayleigh's criterion to 2D imaging:

\vspace{0.1in}
\noindent\textbf{Theorem 2 (Modern Rayleigh's criterion for two-dimensional imaging)}: \emph{
For imaging of incoherent point sources in the subdiffraction limit, the estimation variance of any moment $M_{k\ell}$ with $k+\ell>2$ increases inverse-polynomially as $s$ decreases;
however, the estimation variance of the second moment $M_{20}$, $M_{11}$ and $M_{02}$ are bounded by a constant independent of $s$.}
\vspace{0.1in}

\begin{figure}[htbp] 
\subfigure[\label{fig:2D-dist1}]{\includegraphics[width=8.5cm]{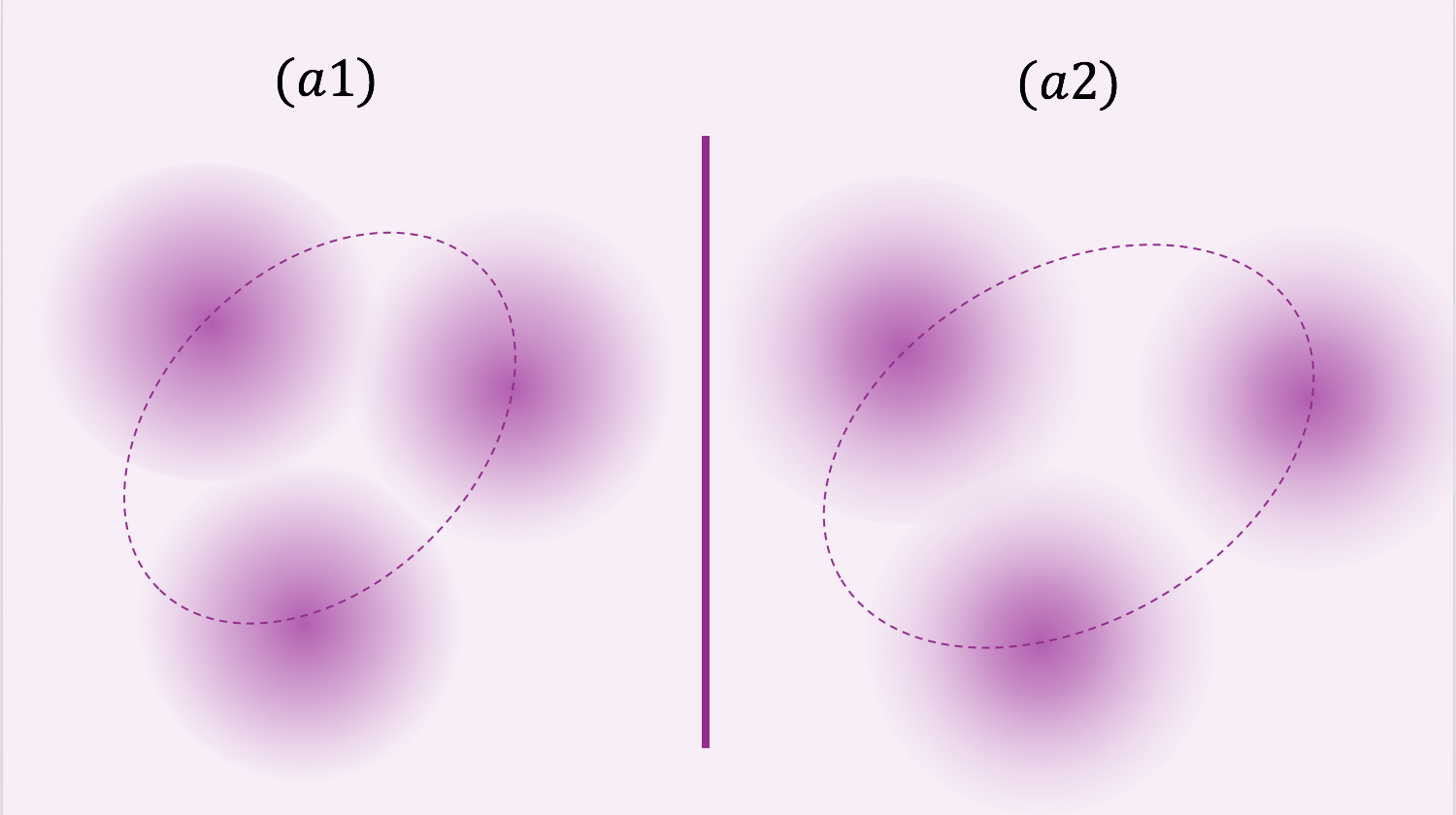}
}
\hspace{0.2cm}
\subfigure[\label{fig:2D-dist2}]{\includegraphics[width=8.5cm]{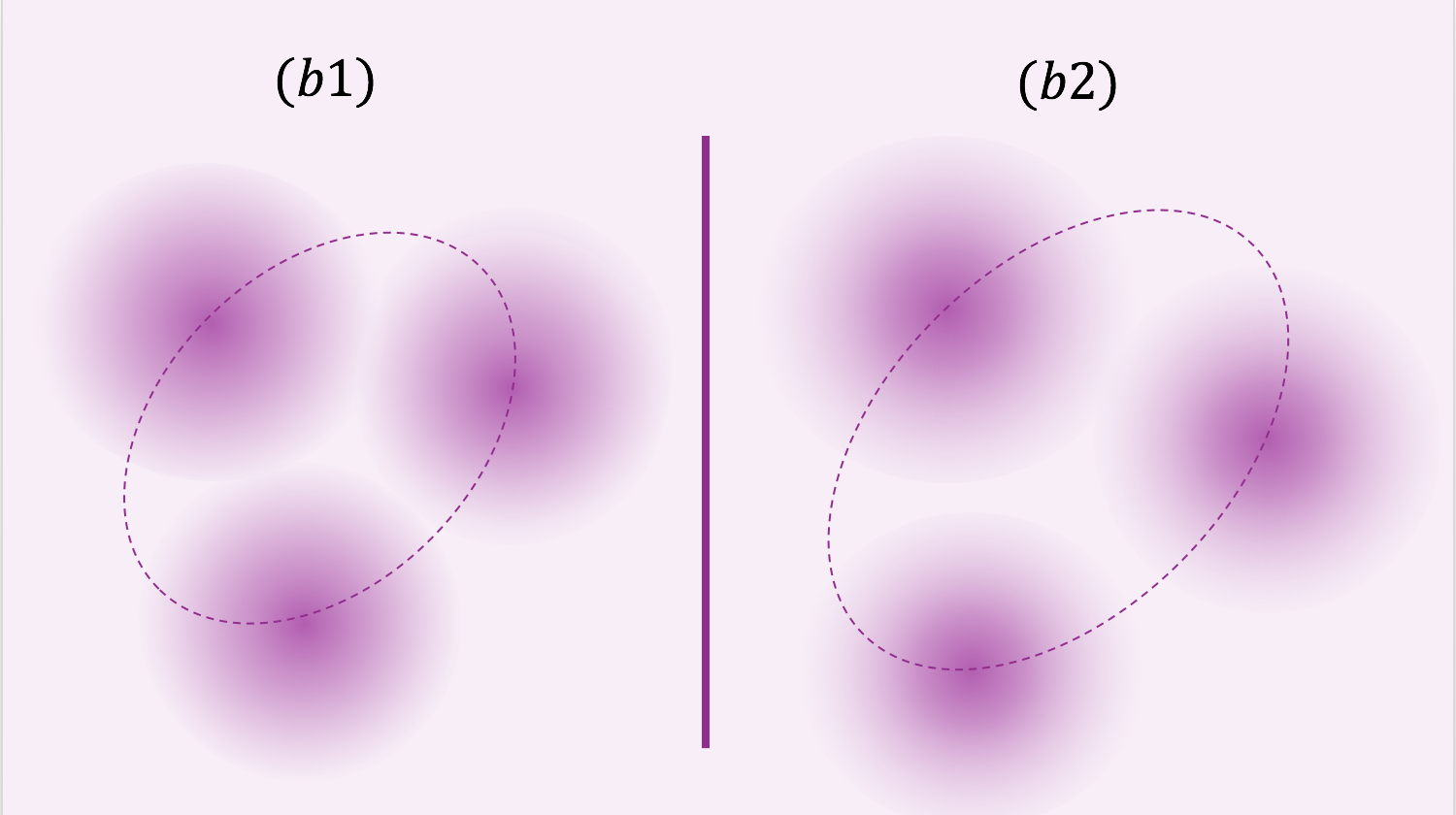}
}
\\
\subfigure[\label{fig:2D-dist3}]{\includegraphics[width=8.5cm]{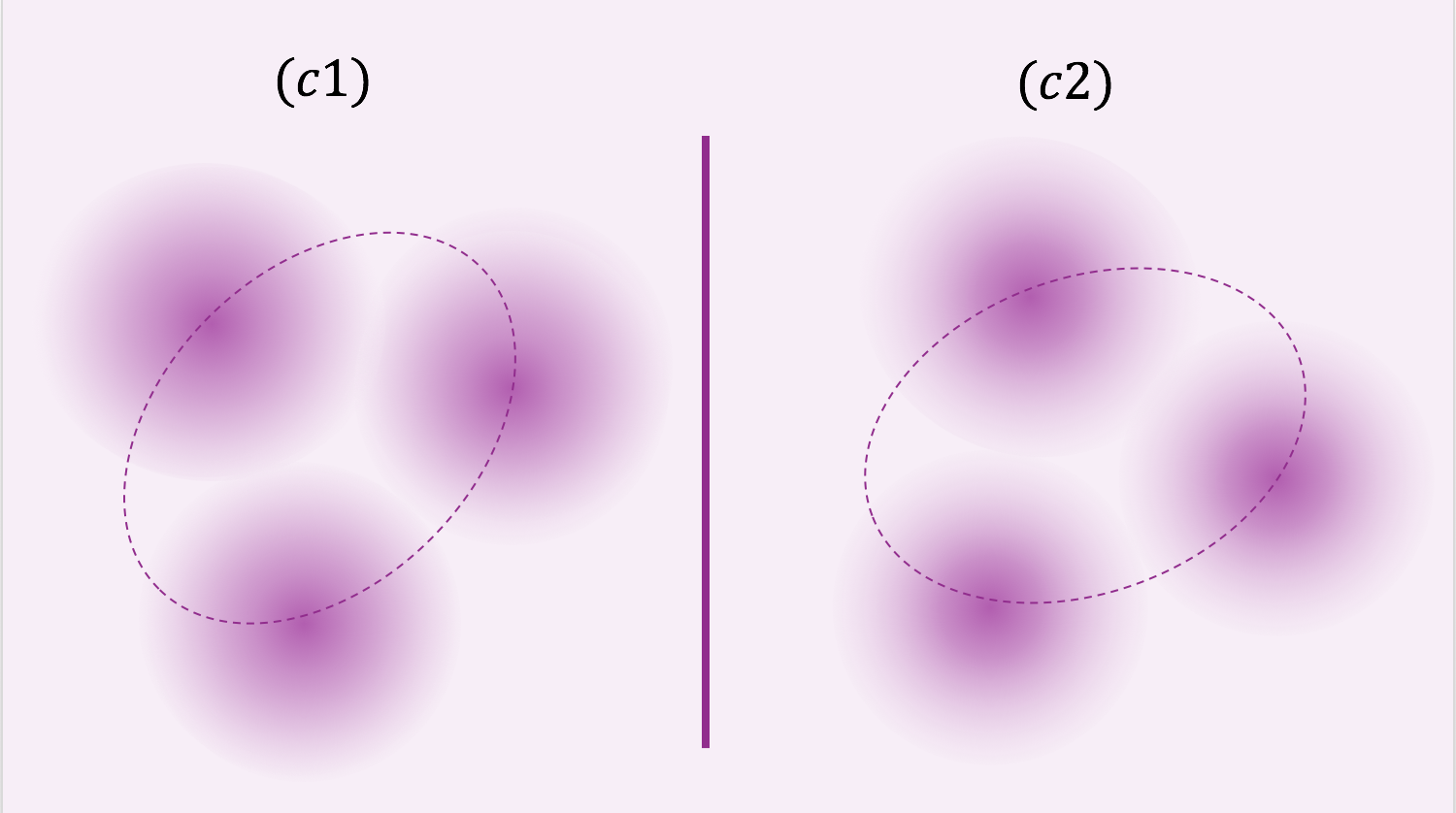}
}
\hspace{0.2cm}
\subfigure[\label{fig:2D-indist}]{\includegraphics[width=8.5cm]{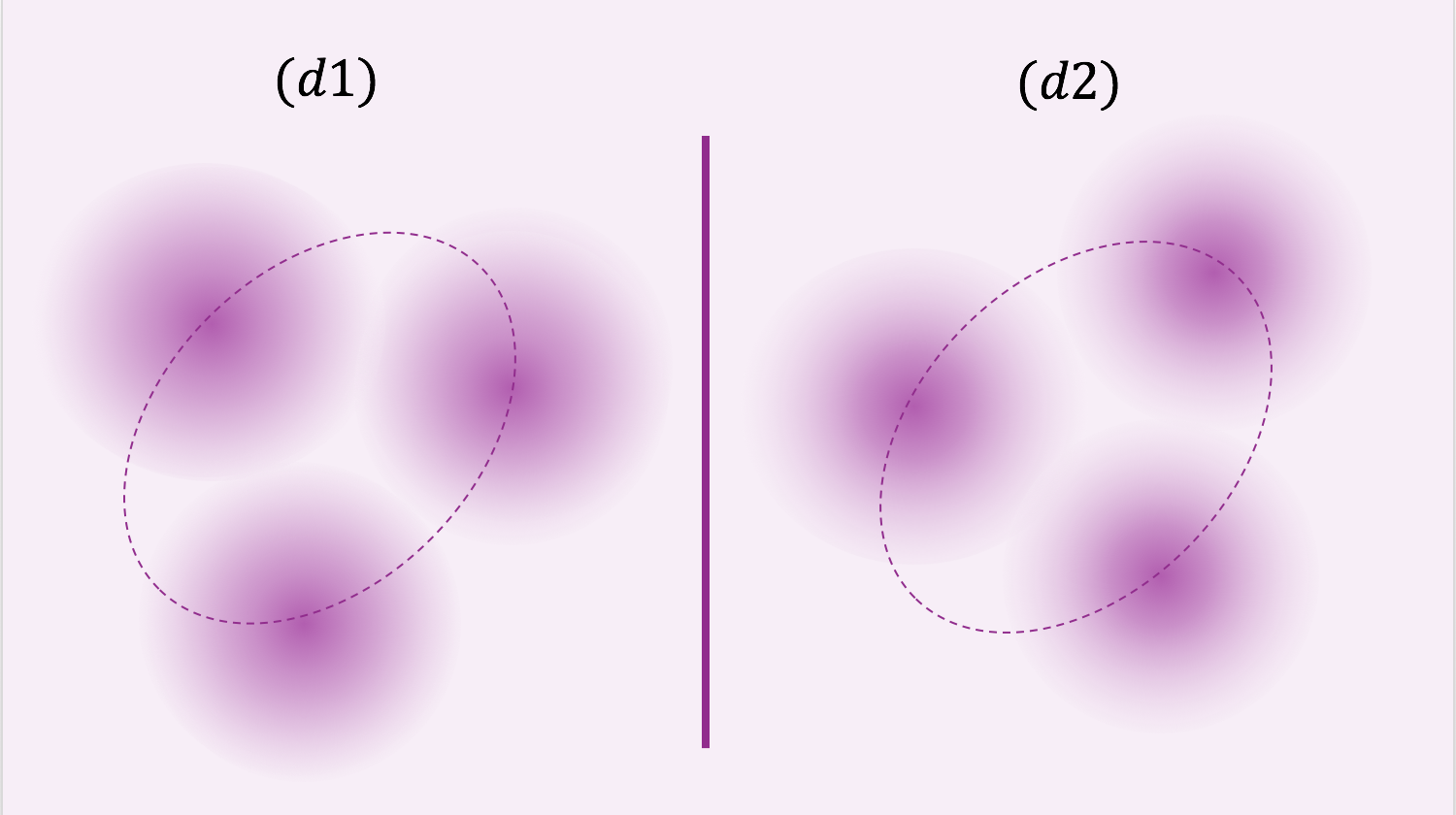}
}
\\
\caption{\label{fig:2D} Three point sources with equal source strengths. Given the values of $(M_{20}, M_{11}, M_{02})$, three points are distributed on an ellipse $\frac{2}{3}(1-\frac{M_{11}^4}{M_{20}^2M_{02}^2}) = \frac{x^2}{M_{20}^2} + \frac{y^2}{M_{02}^2} - \frac{2M_{11}^2 xy}{M_{20}^2M_{02}^2}$. (a)~Images (a1) and (a2) are distinguishable due to different standard deviations along $x$-axis $X = M_{20}$ (b)~Images (b1) and (b2) are distinguishable due to different standard deviations along $y$-axis $Y = M_{02}$. (c)~Images (c1) and (c2) are distinguishable due to different $x$-$y$ correlations $\beta = M_{11}^2/(M_{20}M_{02})$. (d)~Images (d1) and (d2) have the same $(M_{20}, M_{11}, M_{02})$. It is difficult to distinguish them from each other.
}
\end{figure}

A simple schematic illustration above theorem is shown in \autoref{fig:2D}. 
We are now going to find the exact values of FI wrt $M_{20}$, $M_{11}$ and $M_{02}$ and corresponding optimal measurements. 
For simplicity we consider the weak source scenario. For arbitrary source strength, the FIs are still the same and the optimal measurements $E(n)$ should be replaced with $\sum_{k=0}^\infty \frac{1}{k!}(\psi_{{\bar X}}^{\dagger})^kE(n)(\psi_{{\bar X}})^{k}$ because
\begin{multline}
Q_{2}(n)=\sum_{k=0}^{\infty}\frac{\epsilon^{k+1}}{k!(1+\epsilon)^{k+1}}\big(\bra{0}(\psi_{{\bar X}})^{k}\psi_{{\bar X} {\bar Y}}^{(10)}E(n)\psi_{{\bar X} {\bar Y}}^{(10)\dagger}(\psi_{{\bar X}}^{\dagger})^{k}\ket{0} M_{20}^2 \\
+ 2\mathrm{Re}[\bra{0}(\psi_{{\bar X}})^{k}\psi_{{\bar X} {\bar Y}}^{(10)}E(n)\psi_{{\bar X} {\bar Y}}^{(01)\dagger}(\psi_{{\bar X}}^{\dagger})^{k}\ket{0}] M_{11}^2
+ \bra{0}(\psi_{{\bar X}})^{k}\psi_{{\bar X} {\bar Y}}^{(01)}E(n)\psi_{{\bar X} {\bar Y}}^{(01)\dagger}(\psi_{{\bar X}}^{\dagger})^{k}\ket{0} M_{02}^2\big),
\end{multline}
which is a generalization of \eqref{eq:q2} from 1D to 2D.

Suppose $[\psi_{{\bar X} {\bar Y}}^{(10)\dagger},\psi_{{\bar X} {\bar Y}}]=[\psi_{{\bar X} {\bar Y}}^{(01)\dagger},\psi_{{\bar X} {\bar Y}}]=0$ and $\braket{0|\psi_{\bar X \bar Y}^{(10)}\psi_{\bar X \bar Y}^{(01)\dagger}|0} \in \mathbb{R}$. This assumption is satisfied, for example, when the PSF is real. The second order term of $P(n;\{x_j,y_j,\Gamma_j\})$ is
\begin{multline}
Q_2(n) \\ = \epsilon \Big(\bra{0}\psi_{{\bar X} {\bar Y}}^{(10)}E(n)\psi_{{\bar X} {\bar Y}}^{(10)\dagger}\ket{0} M_{20}^2 
+ 2\mathrm{Re}[\bra{0}\psi_{{\bar X} {\bar Y}}^{(10)}E(n)\psi_{{\bar X} {\bar Y}}^{(01)\dagger}\ket{0}] M_{11}^2
+ \bra{0}\psi_{{\bar X} {\bar Y}}^{(01)}E(n)\psi_{{\bar X} {\bar Y}}^{(01)\dagger}\ket{0} M_{02}^2\Big) + O(\epsilon^2).
\end{multline}
We only consider $0$-null measurement outcome $n \in N^{\rm w}_0=\{n|\braket{0|E(n)|0} = \braket{0|\psi_{{\bar X} {\bar Y}}E(n)\psi_{{\bar X} {\bar Y}}^{\dagger}|0}=0,\;\forall k\}$ because for $n \notin N^{\rm w}_0$, the zeroth order term of $P(n;\{x_j,y_j,\Gamma_j\})$ would be positive and does not contribute to the FI as $s\rightarrow 0$. 
Furthermore, we assume $E(n)=\Pi E(n) \Pi$ where $\Pi$ is the projection onto the space $\mathrm{span}\{\psi_{{\bar X} {\bar Y}}^{(10)\dagger}\ket{0},\psi_{{\bar X} {\bar Y}}^{(01)\dagger}\ket{0}\}$ because any component of $E(n)$ perpendicular to it does not 
contribute to $Q_2(n)$ in the first order expansion of $\epsilon$ and consequently only affects the value of the FI in higher order terms of $\epsilon$.

Then we can write every operator as a two-dimensional matrix in basis 
\begin{equation}
\Big\{\ket{e_1} = \frac{1}{\sqrt{2(1+r)}}\bigg(\frac{\psi_{{\bar X} {\bar Y}}^{(10)\dagger}}{\Delta k_x}+\frac{\psi_{{\bar X} {\bar Y}}^{(01)\dagger}}{\Delta k_y}\bigg)\ket{0}, 
\ket{e_2} = \frac{1}{\sqrt{2(1-r)}}\bigg(\frac{\psi_{{\bar X} {\bar Y}}^{(10)\dagger}}{\Delta k_x}-\frac{\psi_{{\bar X} {\bar Y}}^{(01)\dagger}}{\Delta k_y}\bigg)\ket{0}\Big\},
\end{equation}
where $\Delta k_x^2 := \braket{0|\psi_{\bar X \bar Y}^{(10)}\psi_{\bar X \bar Y}^{(10)\dagger}|0} = \int dxdy\; \big|\partial_x\psi_{\mathrm{PSF}}(x,y)\big|^{2}$, $\Delta k_y^2 := \braket{0|\psi_{\bar X \bar Y}^{(01)}\psi_{\bar X \bar Y}^{(01)\dagger}|0}  = \int dxdy\; \big|\partial_y\psi_{\mathrm{PSF}}(x,y)\big|^{2}$ and $r := \braket{0|\psi_{\bar X \bar Y}^{(10)}\psi_{\bar X \bar Y}^{(01)\dagger}|0}/(\Delta k_x \Delta k_y)= \frac{1}{\Delta k_x \Delta k_y}\int dxdy\; \partial_x\psi^*_{\mathrm{PSF}}(x,y)\partial_y\psi_{\mathrm{PSF}}(x,y) \in (-1,1)$. 
Therefore,
\begin{equation}
Q_2(n)  \approx \epsilon \;\trace({E(n) \rho_2}),
\end{equation}
where 
\begin{equation}
\label{eq:rho2}
\rho_2 = \frac{1}{2}\Big( (\Delta k_x^2M_{20}^2+\Delta k_y^2M_{02}^2)(I+r\sigma_z)+2\Delta k_x\Delta k_y M_{11}^2(rI+\sigma_z)+\sqrt{1-r^2}(\Delta k_x^2M_{20}^2-\Delta k_y^2M_{02}^2)\sigma_x\Big).
\end{equation}
Note that $\rho_2$ depends not only on the PSF via $(\Delta k_x,\Delta k_y,r)$ but also on the second moments. The FIM can be then be calculated using $Q_2(n)$ for any specific POVM $\{E(n)\}$.

One way to parametrize the second moments is to define $M_{20} = X^2$, $M_{02} = Y^2$ and $M_{11} = \beta XY$, where $X$, $Y$ is the standard deviation along $x$- and $y$- axis and $\beta$ is the correlation between the distributions along $x$- and $y$- axis. If we approximate the image by a Gaussian distribution $P(x,y) = \frac{1}{2\pi XY\sqrt{1-\beta^2}} \exp (-\frac{1}{2(1-\beta^2)}(x\;y)\v{C}^{-1}(x\;y)^T)$, where the covariance matrix
\begin{equation}
\v{C} 
= \begin{pmatrix}M_{20}^2 & M_{11}^2 \\ M_{11}^2 & M_{02}^2\end{pmatrix}
= \begin{pmatrix}X^2 & \beta XY\\ \beta XY& Y^2\end{pmatrix} 
\end{equation}
the contour lines of $P(x,y)$ will be ellipses described by $\frac{x^2}{X^2}+\frac{y^2}{Y^2}-\frac{2\beta xy}{XY} = 
\mathrm{constant}$. Different distributions can be distinguished from each other if we can precisely estimate the values of $(X,Y,\beta)$.
Another way to parametrize the second moments is to use
\begin{equation}
\v{C} 
= \begin{pmatrix}M_{20}^2 & M_{11}^2 \\ M_{11}^2 & M_{02}^2\end{pmatrix}
= \begin{pmatrix}\cos\theta & -\sin\theta\\ \sin\theta & \cos\theta \end{pmatrix} \begin{pmatrix}\Lambda_1^2 & 0 \\ 0 & \Lambda_2^2 \end{pmatrix} \begin{pmatrix}\cos\theta & \sin\theta\\ -\sin\theta & \cos\theta \end{pmatrix},
\end{equation}
The major and minor length of the ellipses correspond to the square root ${\Lambda_{1,2}}$ of the eigenvalues of $\v{C}$ and the orientation $\theta$ is associated with the direction of its eigenvectors. Estimation wrt $(\Lambda_1,\Lambda_2,\theta)$ is discussed in \appref{app:angle}. 

First let's consider the singular case where $\beta = 1$, $|M_{11}^2| = \sqrt{M_{20}^2M_{02}^2}$ and $\rho_2$ is pure. 
It happens when all points sources are aligned on the same line, e.g. when there are only two point sources \cite{ang2017quantum}. The optimal measurement can be determined by calculating quantum Fisher information matrix (QFIM) wrt $X$ and $Y$:
\begin{equation}
\label{eq:qfi}
\mathcal J_{\mu\nu} = \epsilon\; \trace({\frac{\mathcal L_{\mu}\mathcal L_{\nu}+\mathcal L_{\nu}\mathcal L_{\mu}}{2}\rho_2}),\quad\mu,\nu=X,Y
\end{equation} 
where the Hermitian operator $\mathcal L_\mu$ is the symmetric logarithmic derivative of $\rho_2$ wrt $\mu$ defined via $\partial_\mu\rho_2 = \frac{1}{2} (\mathcal L_\mu\rho_2 + \rho_2\mathcal L_\mu)$ \cite{braunstein1994statistical}. The QFIM derived from \eqref{eq:rho2} is 
\begin{equation}
\mathcal J[X,Y] = 4\epsilon\begin{pmatrix} \Delta k_x^2 & r\Delta k_x \Delta k_y \\ r \Delta k_x \Delta k_y & \Delta k_y^2\end{pmatrix}.
\end{equation} 
The optimal measurement can be chosen to be any rank-one projection onto an orthonormal basis of the real space spanned by $\{\ket{e_1},\ket{e_2}\}$, such as $\{E(n_1) = \ket{e_1}\bra{e_1},\;E(n_2)=\ket{e_2}\bra{e_2}\}$ (the same as 2D-SPADE for Gaussian PSFs \cite{ang2017quantum}) or $\{E(n_1) = \ket{e_+}\bra{e_+},\; E(n_2) = \ket{e_-}\bra{e_-}\}$ where $\ket{e_+}=\frac{1}{\Delta k_x}\psi_{{\bar X} {\bar Y}}^{(10)\dagger}\ket{0}$ and $\ket{e_-}=\frac{1}{\Delta k_y}\psi_{{\bar X} {\bar Y}}^{(01)\dagger}\ket{0}$, 
because they will always satisfy the QFIM-achievable condition $E(n_i) \rho_2^{1/2} = c_{i,\mu} E(n_i)  \mathcal L_{\mu}\rho_2^{1/2}$ for $i=1,2$, $\mu= X,Y$ and some real constant $c_{i,\mu}$. 

For 2D PSF satisifying the following more strict assumption (generalized from \eqref{eq:prop-1D}):
\begin{equation}
\label{eq:prop-2D}
\int_{-\infty}^\infty \Big(\frac{d^{\ell_1}}{dx^{\ell_1}}\frac{d^{\ell_2}}{dy^{\ell_2}}\psi^*_{\mathrm{PSF}}(x,y)\Big)\Big(\frac{d^{\ell_3}}{dx^{\ell_3}}\frac{d^{\ell_4}}{dy^{\ell_4}}\psi_{\mathrm{PSF}}(x,y)\Big) dxdy= 0, \text{ when }\abs{\ell_1-\ell_3}=1\text{ or }\abs{\ell_2-\ell_4}=1,
\end{equation}
the FIM and corresponding measurement can be obtained in a simpler form (otherwise, the FIM can have off-diagonal terms). \eqref{eq:prop-2D} is still quite general. When $\psi_{\mathrm{PSF}}(x,y)$ is separable, i.e. $\psi_{\mathrm{PSF}}(x,y) = \psi_{1,\mathrm{PSF}}(x)\psi_{2,\mathrm{PSF}}(y)$, \eqref{eq:prop-2D} is automatically satisifed when $\psi_{1,\mathrm{PSF}}(x)$ and $\psi_{2,\mathrm{PSF}}(y)$ satisify \eqref{eq:prop-1D}, e.g. $\psi_{\mathrm{PSF}}(x,y) \propto e^{-(x^2+y^2)/4\sigma^2}$ or $\psi_{\mathrm{PSF}}(x,y) \propto e^{ik(x^2+y^2)/2z} {\rm sinc}(x/\sigma_1){\rm sinc}(y/\sigma_2)$. When $\psi_{\mathrm{PSF}}(x,y)$ is a circularly-symmetric function, i.e. $\psi_{\mathrm{PSF}}(x,y) = \psi_{\mathrm{PSF}}(\sqrt{x^2+y^2})$, \eqref{eq:prop-2D} is also true, e.g. $\psi_{\mathrm{PSF}}(x,y) \propto e^{ik(x^2+y^2)/2z}\frac{J_1\big(\sqrt{x^2+y^2}/\sigma\big)}{\sqrt{x^2+y^2}}$ ($J_1(\cdot)$ is the first order Bessel function of the first kind). We assume from now on that \eqref{eq:prop-2D} is satisfied for any 2D PSF. In this case, $r = 0$.

Note that if the projection $\{\Pi E(n)\Pi \}$ of measurements  $\{E(n)\}$ onto the complex space spanned by $\{\ket{e_1},\ket{e_2}\}$ is optimal, $\{E(n)\}$ is also optimal. In particular, 
when $\psi_{\mathrm{PSF}}(x,y)$ is circularly symmetric, any measurement satisfying $\{\Pi E(n_1) \Pi= \ket{e_+}\bra{e_+},\; \Pi E(n_2) \Pi= \ket{e_-}\bra{e_-}\}$ is optimal
, including $\{E(n_1) = \frac{(I - \mathcal P_1)(I + \mathcal P_2)}{4},\; E(n_2) = \frac{(I + \mathcal P_1)(I - \mathcal P_2)}{4}\}$ where the parity operators $\mathcal P_{1(2)}$ satisfies $\mathcal P_{1(2)} f(x,y) = f(- x, y)\; (f(x,-y))$ (the same as 2D-SLIVER \cite{ang2017quantum}). This type of measurement does not depend on the specific expressions of PSFs. In fact, any measurement $E(n)= \sum_{\mu\nu=+,-} m_{\mu,\nu}\ket{e_\mu}\bra{e_\nu}$ can be transformed into a PSF-independent version by replacing $\ket{e_+}\bra{e_+}$ with $\frac{(I - \mathcal P_1)(I + \mathcal P_2)}{4}$, $\ket{e_+}\bra{e_-}$ with $\frac{(I - \mathcal P_1)(I + \mathcal P_2)}{4}\mathcal S_{12}$, $\ket{e_-}\bra{e_+}$ with $\mathcal S_{12}\frac{(I - \mathcal P_1)(I + \mathcal P_2)}{4}$ and $\ket{e_-}\bra{e_-}$ with $\frac{(I + \mathcal P_1)(I - \mathcal P_2)}{4}$ where $\mathcal S_{12} f(x,y) = f(y,x)$.

When $M_{20}$, $M_{11}$ and $M_{02}$ are indepedent parameters, $\beta < 1$, $|M_{11}^2| < M_{20}^2 M_{02}^2$ and $\rho_2$ is a mixed state. 
The QFIM wrt $(X,Y,\beta)$ is
\begin{equation}
\mathcal J[X,Y,\beta] = 4\epsilon
\begin{pmatrix} 
\Delta k_x^2 & 0 & 0\\ 
0 & \Delta k_y^2 & 0\\
0 & 0 & \frac{\Delta k_x^2\Delta k_y^2 X^2 Y^2}{(\Delta k_x^2 X^2 + \Delta k_y^2 Y^2)(1-\beta^2)}\\
\end{pmatrix}.
\end{equation} 
However, the QFIM is not simultaneously achievable for $(X,Y,\beta)$, meaning the quantum Cram\'er-Rao bound $\Sigma \succeq \mathcal{J}^{-1}$ it not attainable. The optimal measurement for $(X,Y)$ is $\{\Pi E(n_1) \Pi= \ket{e_+}\bra{e_+},
\; \Pi E(n_2) \Pi= 
\ket{e_-}\bra{e_-}\}$
and the optimal measurement for $\beta$ is $\{\Pi E(n_1) \Pi= \ket{e'_1}\bra{e'_1},\; \Pi E(n_2) \Pi= \ket{e'_2}\bra{e'_2}\}$, where 
\begin{align}
\ket{e'_1} &= \cos\theta' \ket{e_1} + \sin\theta' \ket{e_2},\\ 
\ket{e'_2} &= -\sin\theta' \ket{e_1} + \cos\theta' \ket{e_2},
\end{align}
and $\theta' =\frac{1}{2} \tan^{-1} \big(\frac{\beta(X^2-Y^2)}{2 X Y}\big)$. 
We note that when $\beta = 0$, the optimal measurement basis for $(X,Y)$ and $\beta$ are mutually unbiased. In fact, any three parameters characterizing $\rho_2$ can never be measured simultaneously using projection-valued measurement (PVM) because $\rho_2$ is only rank two. In practice, we can switch between different types of measurements during the measurement process. 
The resultant FIM will be the average of FIMs wrt each measurement.

\section{Estimation of all moments in the subdiffraction limit}
\label{sec:all-moments}

Even though the information of normalized moments with an order higher than two is jeopardized in the subdiffraction limit, it is worth figuring out the maximum FI achievable and the optimal measurement corresponding to it as one may still need to measure the high-order normalized moments even when the FI is low and the estimation cost is expensive. In this section, we will assume all moments are inpedendent variables and we only consider weak source scenario here. Generalization to sources with arbitrary strengths is contained in the \appref{app:generalization}. Ref.~\cite{tsang2017subdiffraction} contains a detailed discussion on the special case where the source is weak and the PSF is Gaussian, but the optimality was not proved there.

\eqref{eq:prop-1D} and \eqref{eq:prop-2D} are satisfied for PSFs in this section and the main result in this section can be summarized in this theorem:

\vspace{0.1in}
\noindent\textbf{Theorem 3 (Optimal precision scaling wrt all moments)}: \emph{
For any moment $M_{K}$ ($M_{L\,(K-L)}$) in 1D (2D) imaging with arbitrary source strengths, the estimation variance is at least $O(s^{2-K})$ when $K$ is even and $O(s^{1-K})$ when $K$ is odd. 
}
\vspace{0.1in}

For directly imaging, the denominator in \eqref{eq:fi} and \eqref{eq:fi-2} are always $O(1)$ and the Fisher information wrt $M_{K}$ or $M_{L\,(K-L)}$ will be $O(s^{2K-2})$ which is $O(s^{K})$ ($O(s^{K-1})$) times smaller than the maximum FI $O(s^{K-2})$ ($O(s^{K-1})$) for even (odd) moments we obtain here. 

For simplicity, let's first look at the one-dimensional case with weak sources $(\epsilon \ll 1)$, (The analysis for arbitrary source strengths is detailed in \appref{app:generalization}.) According to \eqref{eq:order} and \eqref{eq:fi}, the lowest power of $s$ $\mathcal{F}_{k\ell}$ can attain is $\max\{k,\ell\}-2$ if and only if there is an $E(n)$ such that $P(n;\{x_j,\Gamma_j\})$ is zero until the $\min\{k,\ell\}$-th order of $s$. However, this condition is not necessarily satisfiable for each moments. 

In order for $p_0(n) = \bra{0}\psi_{{\bar X}} E(n) \psi_{{\bar X}}^\dagger\ket{0} = 0$, $E(n)$ has to be orthogonal to $\psi_{{\bar X}}^\dagger\ket{0}$ ($\psi_{{\bar X}}^\dagger\ket{0}$ is not in the support of $E(n)$). Similarly, according to \eqref{eq:q2}, in order for $Q_2(n)$ (up to the first order of $\epsilon$) to be zero, $E(n)$ has to be orthogonal to $\psi_{{\bar X}}^{(1)\dagger}\ket{0}$. We define $\ell$-null measurement outcomes
\begin{equation}
N^{\rm w}_{\ell} = \{n|\bra{0}E(n)\ket{0} = \bra{0}\psi_{{\bar X}}^{(k)}E(n)\psi_{{\bar X}}^{(k)\dagger}\ket{0} = 0,\forall k\leq \ell\},
\end{equation} 
and we have $N^{\rm w}_{\ell} \subseteq N^{\rm w}_{\ell-1}$ for all $\ell$, that is, $\ell$-null measurement is $(\ell-1)$-null. Then for all $\ell\geq0$, $Q_{2\ell}=O(\epsilon^2)$ requires $n\in N^{\rm w}_{\ell}$. Suppose $n \in N^{\rm w}_{\ell-1}$, then
\begin{equation}
Q_{k}(n;\{M_{k'},k'\leq k\}) = O(\epsilon^2), \;\;\forall k\leq 2\ell-1
\end{equation}
and 
\begin{equation}
\label{eq:q2-even}
Q_{2\ell}(n;\{M_k,k\leq 2\ell\}) = \frac{\epsilon}{\ell!^2}  \braket{0|\psi_{{\bar X}}^{(\ell)}E(n)\psi_{{\bar X}}^{(\ell)\dagger}|0}(M_{2\ell})^{2\ell} + O(\epsilon^2).
\end{equation}

We assume derivatives of the PSF $\{\partial^{k}_{{\bar X}}\psi_{\mathrm{PSF}}(x-{\bar X}),k\geq0\}$ form a linear independent subset in $L^2(\mathbb{C})$. 
An orthonormal set $\{b^{(k)}(x),k\geq0\}$ can be generated via Gram-Schmidt process such that $b^{(\ell)}(x)$ is orthogonal to every $\partial^{k}_{{\bar X}}\psi_{\mathrm{PSF}}(x-{\bar X})$ with $k\leq \ell-1$ and 
\begin{equation}
q_\ell := \frac{1}{\ell!}\int b^{(\ell)*}(x) \partial^{\ell}_{{\bar X}}\psi_{\mathrm{PSF}}(x-{\bar X}) dx \in \mathbb{R}.
\end{equation} 
For example, when the PSF is Gaussian, $\{b^{(k)}(x),k\geq 0\}$ are the Hermite-Gaussian modes; when the PSF is a sinc function, $\{b^{(k)}(x),k\geq 0\}$ are the spherical Bessel functions of the first kind. We also notice that, according to \eqref{eq:prop-1D}, $\mathrm{span}\{b^{(k)}(x),k\text{ is even}\} = \mathrm{span}\{\partial^{k}_{{\bar X}}\psi_{\mathrm{PSF}}(x-{\bar X}),k\text{ is even}\}$, $\mathrm{span}\{b^{(k)}(x),k\text{ is odd}\} = \mathrm{span}\{\partial^{k}_{{\bar X}}\psi_{\mathrm{PSF}}(x-{\bar X}),k\text{ is odd}\}$ and they are orthogonal subspaces. 

Then $\mathcal F_{2\ell\,2\ell}$ is maximized when $b_{{\bar X}}^{(\ell)\dagger}|0\rangle$ is an eigenstate of $E(n)$ with an eigenvalue equal to one. The resultant FI is
\begin{equation}
\label{eq:1D-even}
\max_{\{E(n)\}}\mathcal F_{2\ell\,2\ell} \approx \epsilon q_\ell^2 (2\ell)^2 (M_{2\ell})^{2\ell-2}  = O(s^{2\ell -2}).
\end{equation}
For example when $\ell = 1$, $b^{(1)}(x) = \frac{1}{\Delta k}\partial_{{\bar X}}\psi_{\mathrm{PSF}}(x-{\bar X})$ and \eqref{eq:1D-even} gives \eqref{eq:fi2-weak}. 

We can show that it is possible for the FI to attain the lowest power of $s$ (the highest precision) for even moments. To be specific, if $k=2\ell$ is even, by projecting quantum states on the image plane onto basis $\{ b_{{\bar X}}^{(\ell)\dagger}|0\rangle,\;\ell\geq 0\}$ ($b_{{\bar X}}^{(\ell)\dagger} = \int dx b^{(\ell)}(x) a_x^\dagger$), $\mathcal F_{kk}$ is maximized and proportional to the $(2\ell-2)$-th power of $s$, as indicated in \eqref{eq:fi}. 
Moreover, according to the Cram\'er-Rao bound (\eqref{eq:CRbound}),
\begin{equation}
\label{eq:CRbound-even}
\Sigma_{2\ell\,2\ell} \geq (\mathcal{F}^{-1})_{2\ell\,2\ell} \geq (\mathcal F_{2\ell\,2\ell})^{-1}.
\end{equation}
The estimation precision of $M_{2\ell}$ is lower bounded by the value of $(\mathcal F_{2\ell\,2\ell})^{-1}$. Meanwhile, the choice of measurement basis $\{ b_{{\bar X}}^{(\ell)\dagger}|0\rangle,\;\ell\geq 0\}$ not only minimizes the value of $(\mathcal F_{2\ell\,2\ell})^{-1}$ but also makes $\mathcal F$ diagonal, which means that the second equality in \eqref{eq:CRbound-even} holds true. Therefore, we conclude that $\{ b_{{\bar X}}^{(\ell)\dagger}|0\rangle,\;\ell\geq 0\}$ is an optimal basis for estimation of even moments for weak incoherent sources. Note that $\{ b_{{\bar X}}^{(\ell)\dagger}|0\rangle,\;\ell\geq 0\}$ may not be a complete basis, but any POVM is optimal as long as it contains projections onto them and other terms $E(n)$ contained in $\{E(n)\}$ is irrelevant because they do not affect the FIM in the lowest order approximation. We do not write out the irrelevant part of POVM in our discussion.

For odd moments, however, the above arguments do not apply. If we require $n \in N^{\rm w}_{\ell}$ to satisfy 
\begin{equation}
\label{eq:oddcondition}
Q_{2\ell}(n;\{M_k,k\leq 2\ell\}) = O(\epsilon^2), 
\end{equation}
then $E(n)$ is not supported by $\psi_{\bar X}^{(k)\dagger}\ket{0}$ for all $k\leq \ell$. Consequently, we have
\begin{equation}
Q_{2\ell+1}(n;\{M_k,k\leq 2\ell+1\}) = \frac{2\epsilon}{\ell!(\ell+1)!} \mathrm{Re}[\braket{0|\psi_{{\bar X}}^{(\ell)}E(n)\psi_{{\bar X}}^{(\ell+1)\dagger}|0}] (M_{2\ell+1})^{2\ell+1} + O(\epsilon^2) = O(\epsilon^2), 
\end{equation}
which implies negligible contribution from weak sources. 

Therefore, in order to take odd moments into account, we need to relax \eqref{eq:oddcondition} by choosing 
$n \in N^{\rm w}_{\ell-1}\backslash N^{\rm w}_{\ell+1}$ to keep the $O(\epsilon)$ term in $Q_{2\ell+1}(n)$.
The coefficient of $(M_{2\ell+1})^{2\ell+1}$ can be non-zero when $E(n)$ is supported by both $\psi_{{\bar X}}^{(\ell+1)\dagger}|0\rangle$ and $\psi_{{\bar X}}^{(\ell)\dagger}|0\rangle$. Meanwhile,
\begin{equation}
Q_{2\ell}(n;\{M_k,k\leq 2\ell\}) = \frac{\epsilon}{\ell!^2}\braket{0|\psi_{{\bar X}}^{(\ell)}E(n)\psi_{{\bar X}}^{(\ell)\dagger}|0} (M_{2\ell})^{2\ell} + O(\epsilon^2)
\end{equation}
would be non-zero at $O(\epsilon)$ too. In the subdiffraction limit ($s \rightarrow 0$), the denominator in \eqref{eq:fi} is dominated by $Q_{2\ell}$ when $n \in N^{\rm w}_{\ell-1}\backslash N^{\rm w}_{\ell+1}$. 
As shown in \appref{app:odd-moments}, we can maximize $\mathcal F_{2\ell+1\,2\ell+1}$ and in the meantime make the estimation of odd moments independent from the estimation of even moments (by letting $\mathcal F_{2\ell+1\,2\ell} = \mathcal F_{2\ell\,2\ell+1} = O(s^{2\ell})$). Then analogous to \eqref{eq:CRbound-even}, $\mathcal F_{2\ell+1\,2\ell+1}$ fully characterizes the estimation precision of $M_{2\ell+1}$. It is maximized when $E(n)$ are projections onto $\{\frac{b_{{\bar X}}^{(\ell)\dagger} \pm b_{{\bar X}}^{(\ell+1)\dagger}}{\sqrt 2}|0\rangle\}$.
Up to the lowest order of $s$ and $\epsilon$,
\begin{equation}
\label{eq:1D-odd}
\begin{split}
\max_{\{E(n)\}}\mathcal{F}_{2\ell+1\,2\ell+1} 
&\approx 4\epsilon q_{\ell+1}^2 (2\ell+1)^2\frac{(M_{2\ell+1})^{4\ell}}{(M_{2\ell})^{2\ell}} = O(s^{2\ell}).
\end{split}
\end{equation}
In the meantime, we can also calculate $\mathcal F_{2\ell\,2\ell}$ which is exactly its optimal value as in \eqref{eq:1D-even}. Therefore, $\{\frac{b_{{\bar X}}^{(\ell)\dagger} \pm b_{{\bar X}}^{(\ell+1)\dagger}}{\sqrt 2}|0\rangle\}$ achieves the optimal precision for both $M_{2\ell}$ and $M_{2\ell+1}$ simultaneously. 

To conclude, we can use the following two subsets of measurement basis
: $B^{\rm w}_1 = \{\frac{b_{{\bar X}}^{(\ell)\dagger} \pm b_{{\bar X}}^{(\ell+1)\dagger}}{\sqrt 2}|0\rangle\;,\ell\text{ is even}\}$ and $B^{\rm w}_2 = \{\frac{b_{{\bar X}}^{(\ell)\dagger} \pm b_{{\bar X}}^{(\ell+1)\dagger}}{\sqrt 2}|0\rangle\;,\ell\text{ is odd}\}$ (divided into two subsets so that they don't overlap) to estimate $\{M_k|k=4k' \text{ or } 4k'+1,k \geq 1\}$ and $\{M_k|k= 4k'+2 \text{ or } 4k'+3,k' \geq 0\}$, respectively. Each moment can be measured with the optimal precision and independently from other moments (the FIM is diagonal). However, each one of $B^{\rm w}_{1,2}$ can only extract half of the whole moment information: 
$B^{\rm w}_1$ estimates moments with orders equal to multiples of 4 plus 0 or 1; $B^{\rm w}_2$ estimates moments with orders equal to multiples of 4 plus 2 or 3. If one only needs to estimate even moments, $B^{\rm w}_0 = \{b_{{\bar X}}^{(\ell)\dagger}|0\rangle\;,\ell\geq 0\}$ is optimal.

Now let's consider the two-dimensional case. Similar to the one-dimensional case, we define 
\begin{equation}
N^{\rm w}_K = \{\bra{0}E(n)\ket{0} =  \bra{0}\psi^{(k\ell)}_{{\bar X}}E(n)\psi^{(k\ell)\dagger}_{{\bar X}}\ket{0} = 0, \;\forall k,\ell,\;\text{s.t.}\;0 \leq k+\ell \leq K\}.
\end{equation} 
Suppose $n \in N^{\rm w}_{K-1}$, the $O(s^{2K})$ term in $P(n;\{x_j,\Gamma_j\})$ would be 
\begin{multline}
\label{eq:q2-even-2D}
Q_{2K}(n;\{M_{k\ell},k+\ell\leq 2K\}) = \\ \sum_{\ell,\ell'=0}^{K} \frac{\epsilon}{\ell!\ell'!(K-\ell)!(K-\ell')!} \braket{0|\psi_{{\bar X} {\bar Y}}^{(\ell\,K-\ell)}E(n)\psi_{{\bar X} {\bar Y}}^{(\ell'\,K-\ell')\dagger}|0} (M_{(\ell+\ell')(2K-\ell-\ell')})^{2K} + O(\epsilon^2).
\end{multline}
$Q_{2K}$ is derived from Taylor expansion of \eqref{eq:prob}. We notice that $Q_{2K}$ can be written as $\mathbb{E}[\bra{\Psi_K}E(n)\ket{\Psi_K}]$ for some unnormalized state $\ket{\Psi_K}$. Hence $Q_{2K}$ is always non-negative and is equal to zero (up to the first order of $\epsilon$) if and only if $n \in N^{\rm w}_K$. Based on the method of induction, we conclude that $Q_{2K} = O(\epsilon^2)$ if and only if $n \in N^{\rm w}_K$. Therefore, by choosing proper measurement basis for $n\in N^{\rm w}_{K-1}\backslash N^{\rm w}_{K}$, one can estimate $M_{L\,2K-L}$ with an FI up to $O(s^{2K-2})$ for all $0 \leq L \leq 2K$. In general, the optimal measurement basis 
depends on the value of each moments. 

For $M_{L\,2K+1-L}$, consider the $O(s^{2K+1})$ term in $P(n;\{x_j,\Gamma_j\})$:
\begin{multline}
\label{eq:q2-odd-2D}
Q_{2K+1}(n;\{M_{k \ell},k+\ell \leq 2K+1\}) = \\ \sum_{\ell,\ell'=0}^{K} \frac{2\epsilon}{\ell!\ell'!(K-\ell)!(K+1-\ell')!} \mathrm{Re}[\braket{0|\psi_{{\bar X} {\bar Y}}^{(\ell\,K-\ell)}E(n)\psi_{{\bar X} {\bar Y}}^{(\ell'\,K+1-\ell')\dagger}|0}] (M_{(\ell+\ell')(2K+1-\ell-\ell')})^{2K+1} + O(\epsilon^2).
\end{multline}
Clearly, if $n \in N^{\rm w}_{K}$, $Q_{2K+1}(n;\{M_{k \ell},k+\ell \leq 2K+1\})$ = 0. Therefore we should focus on measurement $E(n)$ such that $n \in N^{\rm w}_{K-1}\backslash N^{\rm w}_{K+1}$. Similar to 1D imaging, the optimal scaling we can obtained for $M_{L\,2K+1-L}$ 
is $O(s^{2K})$.

Again we assume derivatives of the PSF $\{\partial^{k}_{{\bar X}}\partial^{\ell}_{{\bar Y}}\psi_{\mathrm{PSF}}(x-{\bar X},y-{\bar Y}),\,k,\ell\geq0\}$ form a linear independent subset in $L^2(\mathbb{C})$. 
An orthonormal set $\{b^{(k\ell)}(x),k\geq0\}$ can be generated such that $b^{(k\ell)}(x)$ is orthogonal to every $\partial^{k'}_{{\bar X}}\partial^{\ell'}_{{\bar Y}}\psi_{\mathrm{PSF}}(x-{\bar X},y-{\bar Y})$ with $k+\ell \leq k+\ell$, $(k,\ell)\neq (k',\ell')$ and
\begin{equation}
q_{k\ell} := \frac{1}{k!\ell!}\int b^{(k\ell)*}(x) \partial^{k}_{{\bar X}}\partial^{\ell}_{{\bar Y}}\psi_{\mathrm{PSF}}(x-{\bar X},y-{\bar Y}) dxdy \in \mathbb{R}.
\end{equation} 
Suppose $\psi_{\mathrm{PSF}}(x,y)$ is separable and $\psi_{\mathrm{PSF}}(x,y) = \psi_{1,\mathrm{PSF}}(x)\psi_{2,\mathrm{PSF}}(y)$. One can generate two orthonormal sets $\{b^{(k)}_{1(2)}(x), \forall k \geq 0\}$ via Gram-Schmidt process from the derivatives of $\psi_{1(2),\mathrm{PSF}}(x)$ as in 1D imaging. Then we have $b^{(k\ell)}(x) = b_1^{(k)}(x)b_2^{(\ell)}(x)$.

Similar to 1D imaging, one can project $\rho$ onto these basis to extract information of moments (see \autoref{tb:2D}) and achieve the optimal scaling of $s$ (but not necessarily the optimal coefficients). As before, one type of measurement can only estimate part of all the moments ($1/4$ to be specific) and by combining different types of measurements one can grasp information of all moments. In practice, combining $\{B^{\rm w}_{i}\}_{i=1}^6$ will be enough to extract all the information of moments from $\rho$. 
For further justifications and calculations of FIs see \appref{app:2D}. 

\begin{table}[ht]
\begin{center}
\begin{tabular}{ |c|c|c|c| } 
\hline
Types of measurement & Measurement basis & $L$ & Moments estimated \\
\hline
$B^{\rm w}_0$ & $b_{{\bar X}  {\bar Y}}^{(L\,K-L)^\dagger}\ket{0}$ & $\mathbb{N}$ & $M_{2L,2K-2L}$\\
\hline
$B^{\rm w}_1$ & $\frac{1}{\sqrt{2}} (b_{{\bar X}  {\bar Y}}^{(L\,K-L)^\dagger} \pm b_{{\bar X}  {\bar Y}}^{(L+1\,K-L-1)^\dagger})\ket{0}$ & even & $M_{2L+1,2K-2L-1}$,    \\
\cline{1-3}
$B^{\rm w}_2$ & $\frac{1}{\sqrt{2}} (b_{{\bar X}  {\bar Y}}^{(L\,K-L)^\dagger} \pm b_{{\bar X}  {\bar Y}}^{(L+1\,K-L-1)^\dagger})\ket{0}$ & odd & $(q_{L,K-L}^2 (M_{2L,2K-2L})^{2K} + q_{L+1,K-L-1}^2 (M_{2L+2,2K-2L-2})^{2K})^{\frac{1}{2K}}$\\
\hline
$B^{\rm w}_3$ & $\frac{1}{\sqrt{2}} (b_{{\bar X}  {\bar Y}}^{(L\,K-L)^\dagger} \pm b_{{\bar X}  {\bar Y}}^{(L+1\,K-L)^\dagger})\ket{0}$& even & $M_{2L+1,2K-2L}$,\\
\cline{1-3}
$B^{\rm w}_4$ & $\frac{1}{\sqrt{2}} (b_{{\bar X}  {\bar Y}}^{(L\,K-L)^\dagger} \pm b_{{\bar X}  {\bar Y}}^{(L+1\,K-L)^\dagger})\ket{0}$& odd & $M_{2L,2K-2L}$\\
\hline
$B^{\rm w}_5$ & $\frac{1}{\sqrt{2}} (b_{{\bar X}  {\bar Y}}^{(K-L\,L)^\dagger} \pm b_{{\bar X}  {\bar Y}}^{(K-L\,L+1)^\dagger})\ket{0}$ & even & $M_{2K-2L,2L+1}$,\\
\cline{1-3}
$B^{\rm w}_6$ & $\frac{1}{\sqrt{2}} (b_{{\bar X}  {\bar Y}}^{(K-L\,L)^\dagger} \pm b_{{\bar X}  {\bar Y}}^{(K-L\,L+1)^\dagger})\ket{0}$& odd & $M_{2K-2L,2L}$\\
\hline
\end{tabular}
\caption {\label{tb:2D} Measurement basis and corresponding moments in 2D imaging, details in \appref{app:2D}.}
\end{center}
\end{table}
In the case of sources with arbitrary strenghs, we show in \appref{app:generalization} that the same scaling wrt $s$ is still achievable by replacing every $E(n)$ with $\sum_{k=0}\frac{1}{k!}(\psi_{{\bar X}}^\dagger)^kE(n)(\psi_{{\bar X}})^k$ (or $\sum_{k=0}\frac{1}{k!}(\psi_{{\bar X} {\bar Y}}^\dagger)^kE(n)(\psi_{{\bar X} {\bar Y}})^k$ for 2D imaging) which also give the same FIs as in the weak source scenario. However, the coefficient 
may be further improved using other basis, due to the fact that information of high order moments can be obtained by detecting several low order derivative operators simultaneously, which is neglectable when the source is weak. In contrast to estimation of the second moment, when estimating higher order moments, the optimal precision increases superlinearly (instead of linearly) as the source strength grows in the subdiffraction limit. 

\section{\label{sec:conclusion}Conclusion}

We have performed a comprehensive Fisher information analysis on general imaging scenarios in the subdiffraction limit, where the improvement of image resolution is considered difficult due to the positive width of point spread functions. We conclude that, for any incoherence sources, a 1D or 2D image can be precisely estimated up to its second moment and the higher order moments are difficult to estimate in the sense that the error increase inverse-polynomially as the size of image decreases. The imaging situation considered in the paper is quite general where both the number of point sources and source strengths can be arbitrary. The problem of pre-estimation of centroid is also worked out. 

For real point spread functions, we put forward a measurement scheme which provides the optimal Fisher information in the subdiffraction limit. The measurement basis is constructed based on the derivates of the point spread function, which are closely related to moments of an image. The optimal measurement scheme for second moment is discussed in detail. For higher order moments, compared with direct imaging approach, our measurement scheme guarantees at least a quadratic improvement of Fisher information in terms of the scaling wrt the size of the image. The coefficient of Fisher information is also optimal for weak sources, but can be further improved for strong sources. It is not clear, though, which measurement basis is optimal in terms of the exact value of Fisher information for strong sources. 

The generality of our results has a cost though --- the Fisher information is only calculated in the limiting case where the size of the image tends to zero. Direct calculations for a positive size can be difficult and it remains unsolved how to identify the optimal measurement scheme when the size is not too small (in the subdiffraction limit) and also not too large (the point spread function can be viewed as a delta function). Our results, however, is an important theoretical result towards the ultimate resolution limit for incoherent optical imaging. 

\vspace{0.1in}
\emph{Note added.}---Recently, Ref.~\cite{tsang2018quantum} appeared, which directly calculates the quantum Fisher information wrt moments for subdiffraction incoherent optical imaging. This approach can be applied to all types of measurements, without the non-adaptivity restriction in our analysis. Our results on arbitrary source strength, generalization to two-dimensional imaging and optimal scaling achieving measurement, however, are not covered in Ref.~\cite{tsang2018quantum}. 

\section*{ACKNOWLEDGMENTS}
We acknowledge support from the ARL-CDQI (W911NF-15-2-0067), ARO (W911NF-14-1-0011, W911NF-14-1-0563), ARO MURI (W911NF-16-1-0349), AFOSR MURI (FA9550-14-1-0052, FA9550-15-1-0015), NSF (EFMA-1640959), the Alfred P. Sloan Foundation (BR2013-049), and the Packard Foundation (2013-39273).


\bibliographystyle{apsrev}

\newpage

\appendix

\section{Validity of series expansions of probability and FIM}
\label{app:convergence}

In this section, we justify the series expansion of the probability $P(n;\{x_j,\Gamma_j\})$ around its centroid. For simplicity, we only consider weak sources in 1D imaging. For single-photon measurement, 
\begin{equation}
P(n;\{x_j,\Gamma_j\}) = \epsilon \sum_{j=1}^J \gamma_j \braket{0|\psi_j E(n) \psi_j^\dagger|0}.
\end{equation}
We want to know when the following series will converge uniformly to $P(n;\{x_j,\Gamma_j\})$:
\begin{equation}
\label{eq:series}
\sum_{k=0}^\infty P_{k}(n) (M_k)^k \stackrel{?}{=} P(n;\{x_j,\Gamma_j\}), 
\end{equation}
where 
\begin{equation}
P_{k}(n) = \frac{\epsilon}{k!} \frac{\partial^k}{\partial x_j^k} \braket{0|\psi_j E(n) \psi_j^\dagger|0} \big|_{x_j = \bar X}.
\end{equation}
Let the radius of convergence $R = (\limsup_{k\rightarrow\infty} \abs{P_{k}(n)}^{1/k})^{-1}$, then \eqref{eq:series} converges uniformly as long as $s < R$~\cite{radius}. 

Next we show that the radius of convergence $R \geq R_0$ where $R_0$ independent of $E(n)$.
\begin{equation}
\label{eq:radiusbound}
R_0 = \bigg(\sup_\ell \bigg(\frac{\|\psi_{\rm PSF}^{(\ell)}\|}{\ell!}\bigg)^{1/\ell}\bigg)^{-1},
\end{equation}
where $\psi_{\rm PSF}^{(\ell)}$ represents the $\ell$-th order derivative of $f$ and $\|\psi_{\rm PSF}^{(\ell)}\| = \sqrt{\int_{-\infty}^{\infty} |\psi_{\rm PSF}^{(\ell)}(x)|^2 dx}$. Then
\begin{equation}
\begin{split}
R^{-1} &= \limsup_{k\rightarrow\infty} \abs{P_k(n)}^{1/k} 
\leq \limsup_{k\rightarrow\infty} \bigg|\sum_{\ell=0}^k\frac{1}{\ell!(k-\ell)!} \|\psi_{\rm PSF}^{(k-\ell)}\| \| \psi_{\rm PSF}^{(\ell)}\| \bigg|^{1/k} \leq R_0^{-1}.
\end{split}
\end{equation}
Therefore when $s < R_0 \leq R$, \eqref{eq:series} uniformly converges. For example, for a Gaussian PSF
\begin{equation}
\psi_{\rm PSF}(x) = \frac{1}{(2\pi\sigma^2)^{1/4}} \exp\bigg(-\frac{x^2}{4\sigma^2}\bigg).
\end{equation}
From 
\begin{equation}
\int_{-\infty}^{+\infty} e^{x^2} \bigg( \Big(\frac{d}{dx}\Big)^\ell e^{-x^2} \bigg)^2 dx= \sqrt{\pi} \ell! 2^\ell,
\end{equation}
we see that $R_0 \geq \sigma$ from \eqref{eq:radiusbound}. Therefore in the subdiffraction limit ($s \ll \sigma$), the series expansion is always valid. However, things may break down when $s > R_0$ which may happen if $\psi_{\rm PSF}(x)$ has complex sub-wavelength structure. 

When $s < R_0$, the diagonal element of the Fisher information matrix is 
\begin{equation}
\begin{split}
\mathcal F_{kk} &= \sum_{n} \frac{1}{P(n;\{x_j,\Gamma_j\})}\bigg(\frac{\partial P(n;\{x_j,\Gamma_j\})}{\partial M_k}\bigg)^2\\
&= \sum_{n} \frac{\abs{P_k(n) k M_k^{k-1}}^2}{\abs{P_k(n) M_k^k}} \frac{b_k^2}{a_k},\\
\end{split} 
\end{equation}
where we assume
\begin{equation}
\abs{\frac{P(n;\{x_j,\Gamma_j\})}{P_k(n) M_k^k}} = a_k,
\quad \text{~~and~~}\quad
\abs{\frac{\partial P(n;\{x_j,\Gamma_j\})/\partial M_k}{P_k(n) k M_k^{k-1}}} = b_k.
\end{equation}
Suppose $\frac{b_k^2}{a_k} \leq c_k$, we have 
\begin{equation}
\begin{split}
\mathcal F_{kk} &< \sum_{n} \frac{(P_k(n) k \abs{M_k}^{k-1})^2}{|P_k(n)| \abs{M_k}^k} c_k = c_k k^2 \abs{M_k}^{k-2} \sum_{n} \abs{P_k(n)} \\
&= c_k k^2 \abs{M_k}^{k-2} \left(\frac{\epsilon}{k!} \frac{\partial^k}{\partial x_j^k} \braket{0|\psi_j \big(E(N_+) - E(N_-)\big)\psi_j^\dagger|0} \big|_{x_j = \bar X}\right)\\
&\leq 2c_k k^2 \left(\sum_{\ell=0}^k\frac{\epsilon}{\ell!(k-\ell)!} \|\psi_{\rm PSF}^{(k-\ell)}\| \| \psi_{\rm PSF}^{(\ell)}\| \right)\abs{M_k}^{k-2} = O(s^{k-2}), \\
\end{split} 
\end{equation}
where $N_+ = \{n:P_k(n) \geq 0\}$, $N_- = \{n:P_k(n) < 0\}$ and $E(N_\pm) = \sum_{n\in N_\pm} E(n)$. 

The order-of-magnitude analysis above is valid only when 
\begin{equation}
c_k = \abs{\frac{P_k(n) (M_k)^k}{\sum_{k'=0}^\infty P_{k'}(n) (M_{k'})^{k'}}}\left(\sum_{k'=k}^\infty P_{k'}(n) \frac{ \partial (M_{k'})^{k'}}{\partial M_k}\right)^2 \Big\slash \left(P_k(n) k (M_k)^{k-1}\right)^2
\end{equation}
is reasonably small when $s$ is small. We argue that this is usually true for non-adaptive measurements:
\begin{itemize}
\item Consider first the case when $\abs{P_k(n) (M_{k})^k} \gg \abs{\sum_{k'>k}P_{k'}(n) (M_{k'})^{k'}}$, then clearly 
\begin{equation}
c_k \approx \frac{\abs{P_k(n) (M_k)^k}}{\sum_{k'=0}^\infty P_{k'}(n) (M_{k'})^{k'}}\cdot 1 \lessapprox 1.
\end{equation}
\item When $\abs{P_k(n) (M_{k})^k} \ll \abs{\sum_{k'>k}P_{k'}(n) (M_{k'})^{k'}} = O(s^{k+1})$,
\begin{equation}
c_k \approx \frac{O(s^{k+1})}{\abs{P_k(n) k (M_k)^{k-1}}}
\end{equation}
may be large. 
However, the contribution to $\mathcal F_{kk}$
\begin{equation}
\frac{\abs{P_k(n) k M_k^{k-1}}^2}{\abs{P_k(n) M_k^k}} c_k = O(s^{k})
\end{equation}
is negligible.
\item When $\abs{P_k(n) (M_{k})^k} \approx \abs{\sum_{k'>k}P_{k'}(n) (M_{k'})^{k'}}$ and (when $P_{k'}(n) = 0$ for all $k' \leq k$) the first and second terms in 
\begin{equation}
P(n;\{x_j,\Gamma_j\}) = P_k(n) (M_{k})^k + \sum_{k'>k}P_{k'}(n) (M_{k'})^{k'}
\end{equation}
cancel each other out, up to the lowest order of $s$. Above analysis could become invalid. However, it requires a special design of measurement based on prior knowledge of the moments. We exclude this type of adaptive measurement in our discussion. 
\end{itemize}

\section{First three terms in the series expansion of measurement probability~for~arbitrary~incoherent~sources}
\label{app:expansion}

We aim to expand $P(n,\{x_j,\Gamma_j\})$ in series of $O(s^k)$ where $s$ is the size of the image. To do this we replace $\psi^{\dagger}\alpha$ with $\sum_{k=0}^{\infty}\frac{A^{(k)}}{k!}\psi_{{\bar X}}^{(k)\dagger}$ in \eqref{eq:prob}, where $A^{(k)}=\sum_{j=1}^{J}\alpha_{j}(x_{j}-{\bar X})^{k}$ and $\psi_{{\bar X}}^{(k)\dagger}=\frac{d^{k}}{d{\bar X}^{k}}\int dx \psi_{\mathrm{PSF}}(x-{\bar X})a_{x}^{\dagger}.$

First of all, we calculate the value of denominator which gives 
\begin{equation}
\braket{0|e^{\alpha^{\dagger}\psi}e^{\psi^{\dagger}\alpha}|0} = e^{\int dx |\sum_j\alpha_j\psi_{\mathrm{PSF}}(x-x_j)|^2}.
\end{equation}
Therefore, 
\begin{equation}
P(n,\{x_j,\Gamma_j\}) = 
\mathbb{E}[e^{-\int dx |\sum_j\alpha_j\psi_{\mathrm{PSF}}(x-x_j)|^2}\sum_{k=0}^{\infty} \frac{1}{k!^2} \braket{0|(\alpha^\dagger \psi)^k E(n) (\psi^\dagger \alpha)^k|0}].
\end{equation}
The zeroth order term is 
\begin{equation}
\begin{split}
Q_0(n) &= \sum_{k=0}^{\infty} \frac{1}{k!^2} \mathbb{E}[e^{- |A^{(0)}|^2}|A^{(0)}|^{2k}] \braket{0|\psi_{{\bar X}}^k E(n) (\psi_{{\bar X}}^\dagger)^k|0} \\ &= \sum_{k=0}^{\infty} \frac{\epsilon^k}{k!(1+\epsilon)^{k+1}} \braket{0|\psi_{{\bar X}}^k E(n) (\psi_{{\bar X}}^\dagger)^k|0},
\end{split}
\end{equation}
where we use $\mathbb{E}[ e^{- |A^{(0)}|^2}|A^{(0)}|^{2k}] 
= \frac{k!\epsilon^k}{(1+\epsilon)^{k+1}}.$

The first order term is 
\begin{equation}
\begin{split}
Q_1(n) & =  \sum_{k=1}^\infty \frac{1}{k!^2} (2k) \mathbb{E}[ (e^{- |A^{(0)}|^2})(A^{(0)*})^{k-1}A^{(1)*}(A^{(0)})^{k}] \mathrm{Re}[\braket{0|(\psi_{{\bar X}})^{k-1}\psi_{{\bar X}}^{(1)} E(n) (\psi_{{\bar X}}^\dagger)^k|0} ]\\
& = \sum_{k=0}^\infty \frac{2 \epsilon^{k+1}}{k!(1+\epsilon)^{k+2}} \mathrm{Re}[\braket{0|(\psi_{{\bar X}})^{k}\psi_{{\bar X}}^{(1)} E(n) (\psi_{{\bar X}}^\dagger)^{k+1}|0}] M_1,
\end{split}
\end{equation}
where we use $\mathbb{E}[ (e^{- |A^{(0)}|^2})(A^{(0)*})^{k-1}A^{(1)*}(A^{(0)})^{k}] = \frac{k!\epsilon^k M_1}{(1+\epsilon)^{k+1}}$. 

The second order term is 
\begin{equation}
\label{eq:q2-complete}
\begin{split}
Q_2(n) 
&= \sum_{k=0}^{\infty} \frac{1}{k!^2} \mathbb{E}[(- e^{- |A^{(0)}|^2})(\mathrm{Re}[A^{(0)*}A^{(2)}] + |A^{(1)}|^2 )|A^{(0)}|^{2k}] \braket{0|\psi_{{\bar X}}^{(1)}\psi_{{\bar X}}^{(1)^\dagger}|0}\braket{0|(\psi_{{\bar X}})^k E(n) (\psi_{{\bar X}}^\dagger)^k|0}\\
&+ \sum_{k=1}^{\infty} \frac{1}{k!^2}\big( k  \mathbb{E}[ (e^{- |A^{(0)}|^2})(A^{(0)*})^{k-1}A^{(2)*}(A^{(0)})^{k} ]  \mathrm{Re}[\braket{0|(\psi_{{\bar X}})^{k-1}\psi_{{\bar X}}^{(2)} E(n) (\psi_{{\bar X}}^\dagger)^k|0}]
\\& \qquad \qquad + k(k-1) \mathbb{E}[ (e^{- |A^{(0)}|^2})(A^{(0)*})^{k-2}(A^{(1)*})^2(A^{(0)})^{k} ] \mathrm{Re}[\braket{0|(\psi_{{\bar X}})^{k} E(n) (\psi_{{\bar X}}^{(1)\dagger})^2 (\psi_{{\bar X}}^\dagger)^{k-2}|0}]
\\& \qquad \qquad + k^2\mathbb{E}[ (e^{- |A^{(0)}|^2})|A^{(0)}|^{2k-2}|A^{(1)}|^2 ] \braket{0|(\psi_{{\bar X}})^{k-1}\psi_{{\bar X}}^{(1)} E(n) \psi_{{\bar X}}^{(1)\dagger}(\psi_{{\bar X}}^\dagger)^{k-1}|0}\big) \\
&= \sum_{k=0}^\infty \frac{\epsilon^{k+1}}{k!(1+\epsilon)^{k+1}}\big((\braket{0|(\psi_{{\bar X}})^{k}\psi_{{\bar X}}^{(1)} E(n) \psi_{{\bar X}}^{(1)\dagger}(\psi_{{\bar X}}^\dagger)^{k}|0} + \mathrm{Re}[\braket{0|(\psi_{{\bar X}})^{k}\psi_{{\bar X}}^{(2)} E(n) (\psi_{{\bar X}}^\dagger)^{k+1}|0}] \\& \qquad \qquad - (k+2)\braket{0|\psi_{{\bar X}}^{(1)}\psi_{{\bar X}}^{(1)^\dagger}|0}\braket{0|(\psi_{{\bar X}})^k E(n) (\psi_{{\bar X}}^\dagger)^k|0})M_2^2\big)\\
&+\sum_{k=0}^\infty \frac{\epsilon^{k+2}}{k!(1+\epsilon)^{k+3}}\braket{0|(\psi_{{\bar X}})^{k+2} E(n) (\psi_{{\bar X}}^{(1)\dagger})^2 (\psi_{{\bar X}}^\dagger)^{k}|0}M_1^2 + \sum_{k=0}^\infty \frac{\epsilon^{k+1}(k-\epsilon)}{k!(1+\epsilon)^{k+2}}\braket{0|(\psi_{{\bar X}})^{k}\psi_{{\bar X}}^{(1)} E(n) \psi_{{\bar X}}^{(1)\dagger}(\psi_{{\bar X}}^\dagger)^{k}|0}M_1^2,
\end{split}
\end{equation}
where we use $\mathbb{E}[ (e^{- |A^{(0)}|^2})|A^{(0)}|^{2(k-1)}|A^{(1)}|^2] = \frac{(k-1)!\epsilon^k (M_2^2- M_1^2)}{(1+\epsilon)^k} + \frac{k!\epsilon^k M_1^2}{(1+\epsilon)^{k+1}}$, $\mathbb{E}[ (e^{- |A^{(0)}|^2})|A^{(0)}|^{2(k-1)}A^{(2)*}A^{(0)}] = \frac{k!\epsilon^k M_2^2}{(1+\epsilon)^{k+1}}$ and $\mathbb{E}[ (e^{- |A^{(0)}|^2})(A^{(0)*})^{k-2}(A^{(1)*})^2(A^{(0)})^{k} ] = \frac{k!\epsilon^k M_1^2}{(1+\epsilon)^{k+1}}$. Suppose the centroid is accurately known, we have $M_1 = 0$ and $Q_1(n) = 0$. If we define $N_{0}=\{n|Q_{0}(n)=\sum_{k=0}^{\infty}\frac{\epsilon^{k}}{k!(1+\epsilon)^{k+1}}\braket{0|(\psi_{{\bar X}})^{k}E(n)(\psi_{{\bar X}}^{\dagger})^{k}|0}=0\}=\{n|\braket{0|(\psi_{{\bar X}})^{k}E(n)(\psi_{{\bar X}}^{\dagger})^{k}|0}=0,\;\forall k\}$. For $n\in N_0$, $Q_0(n) = Q_1(n) = 0$ and only the first term in $Q_2(n)$ survives, which gives \eqref{eq:q2}. The second term $\mathrm{Re}[\braket{0|(\psi_{{\bar X}})^{k}\psi_{{\bar X}}^{(2)} E(n) (\psi_{{\bar X}}^\dagger)^{k+1}|0}]$ in \eqref{eq:q2-complete} vanishes for $n\in N_0$ because $E(n)$ is Hermitian and non-negative and its eigenstates corresponding to non-vanishing eigenvalues must be orthogonal to $(\psi_{{\bar X}}^{\dagger})^k\ket{0}$ for all $k$.

\section{An alternative way to parametrize second moments in 2D imaging}{}
\label{app:angle}

Here we calculation the optimal FIM wrt $(\Lambda_1,\Lambda_2,\theta)$ as defined in \autoref{sec:2D}. We only consider the situation where $\Delta k_x = \Delta k_y = \Delta k$ and $r = 0$ as the form of FIM becomes quite complicated otherwise and provides no physical intuition. The QFIM wrt $(\Lambda_1,\Lambda_2,\theta)$ calculated from \eqref{eq:rho2} is 
\begin{equation}
\label{eq:qfi-angle}
\mathcal J[\Lambda_1,\Lambda_2,\theta] = 
\begin{pmatrix}
4\epsilon\Delta k^2 & 0 & 0 \\
0 & 4\epsilon\Delta k^2 & 0 \\ 
0 & 0 & \frac{4\epsilon\Delta k^2(\Lambda_1^2 - \Lambda_2^2)^2}{\Lambda_1^2 + \Lambda_2^2}
\end{pmatrix}.
\end{equation}
It is clear from \eqref{eq:qfi-angle} that when $\Lambda_1 = \Lambda_2$, the QFI is zero, which means when the image is circular-uniformly distributed (up to its second moment), we are not able to estimate $\theta$ in the subdiffraction limit.

The corresponding optimal measurements found from the QFIM calculation are 
\begin{equation}
\label{eq:appbasis-1}
\begin{split}
E(n_1) &= (\cos(\theta+\pi/4)\ket{e_1}-\sin(\theta+\pi/4)\ket{e_2})(\cos(\theta+\pi/4)\bra{e_1}-\sin(\theta+\pi/4)\bra{e_2}),\\
E(n_2) &=(\sin(\theta+\pi/4)\ket{e_1}+\cos(\theta+\pi/4)\ket{e_2})(\sin(\theta+\pi/4)\bra{e_1}+\cos(\theta+\pi/4)\bra{e_2}).
\end{split}
\end{equation} for estimation of $(\Lambda_1,\Lambda_2)$ and 
\begin{equation}
\label{eq:appbasis-2}
\begin{split}
E(n_1) &= (\cos\theta\ket{e_1}-\sin\theta\ket{e_2})(\cos\theta\bra{e_1}-\sin\theta\bra{e_2}),\\
E(n_2) &=(\sin\theta\ket{e_1}+\cos\theta\ket{e_2})(\sin\theta\bra{e_1}+\cos\theta\bra{e_2}).
\end{split}
\end{equation} 
for estimation of $\theta$. We note here that \eqref{eq:appbasis-1} and \eqref{eq:appbasis-2} are mutually unbiased.

\section{Optimization of FI wrt odd moments for weak incoherent sources in 1D imaging}
\label{app:odd-moments}

Up to the lowest order of $s$ and $\epsilon$,
\begin{equation}
\begin{split}
\mathcal F_{2\ell+1\,2\ell+1} &\approx \sum_{n \in N^{\rm w}_{\ell-1}\backslash N^{\rm w}_{\ell+1}} \frac{1}{Q_{2\ell}(n;\{M_k,k\leq 2\ell\})}\bigg(\frac{\partial Q_{2\ell+1}(n;\{M_k,k\leq 2\ell+1\})}{\partial M_{2\ell+1}}\bigg)^{2} \\
&= \frac{4(2\ell+1)^2\epsilon}{(\ell+1)!^2} \frac{(M_{2\ell+1})^{4\ell}}{(M_{2\ell})^{4\ell}}  \sum_{n\in N^{\rm w}_{\ell-1}\backslash N^{\rm w}_{\ell+1}} \frac{ (\mathrm{Re}[\bra{0}\psi_{{\bar X}}^{(\ell)} E(n)\psi_{{\bar X}}^{(\ell+1)^\dagger}\ket{0}])^2 }{ \bra{0}\psi_{{\bar X}}^{(\ell)} E(n)\psi_{{\bar X}}^{(\ell)^\dagger}\ket{0}} 
\end{split}
\end{equation}
First we note that, in order to maximize $\mathcal F_{2\ell+1\,2\ell+1}$, we can assume $E(n)$ is a rank-one projector for each $n$, because for any $E(n) = \sum_{i} p_i \ket{\Phi_i}\bra{\Phi_i}$,
\begin{equation}
\frac{ \big(\sum_{i} p_i\mathrm{Re}[\bra{0}\psi_{{\bar X}}^{(\ell)}\ket{\Phi_i}\bra{\Phi_i}\psi_{{\bar X}}^{(\ell+1)^\dagger}\ket{0}]\big)^2 }{\sum_{i} p_i  \bra{0}\psi_{{\bar X}}^{(\ell)} \ket{\Phi_i}\bra{\Phi_i} \psi_{{\bar X}}^{(\ell)^\dagger}\ket{0}} \leq \sum_{i} p_i \frac{ \big(\mathrm{Re}[\bra{0}\psi_{{\bar X}}^{(\ell)}\ket{\Phi_i}\bra{\Phi_i}\psi_{{\bar X}}^{(\ell+1)^\dagger}\ket{0}]\big)^2}{\bra{0}\psi_{{\bar X}}^{(\ell)} \ket{\Phi_i}\bra{\Phi_i} \psi_{{\bar X}}^{(\ell)^\dagger}\ket{0}}
\end{equation}
according to Cauchy-Schwarz inequality. Therefore deviding any POVM into corresponding projective measurements will only increase FI. Furthermore, if $E(n) = \ket{\Phi_n}\bra{\Phi_n}$,
\begin{equation}
\frac{ \big(\mathrm{Re}[\bra{0}\psi_{{\bar X}}^{(\ell)}\ket{\Phi_n}\bra{\Phi_n}\psi_{{\bar X}}^{(\ell+1)^\dagger}\ket{0}]\big)^2}{\bra{0}\psi_{{\bar X}}^{(\ell)} \ket{\Phi_n}\bra{\Phi_n} \psi_{{\bar X}}^{(\ell)^\dagger}\ket{0}}
\leq 
\frac{|\bra{0}\psi_{{\bar X}}^{(\ell)}\ket{\Phi_n}\bra{\Phi_n}\psi_{{\bar X}}^{(\ell+1)^\dagger}\ket{0}|^2}{\bra{0}\psi_{{\bar X}}^{(\ell)} \ket{\Phi_n}\bra{\Phi_n} \psi_{{\bar X}}^{(\ell)^\dagger}\ket{0}} = \bra{0}\psi_{{\bar X}}^{(\ell+1)}\ket{\Phi_n}\bra{\Phi_n}\psi_{{\bar X}}^{(\ell+1)^\dagger}\ket{0}.
\end{equation}
We can, for example, choose the measurement basis to be $\ket{\Phi_\pm} = \frac{b_{{\bar X}}^{(\ell)\dagger} \pm b_{{\bar X}}^{(\ell+1)\dagger}}{\sqrt{2}}\ket{0}$ (other real superposition of $b_{\bar X}^{(\ell)\dagger}\ket{0}$ and $b_{\bar X}^{(\ell+1)\dagger}\ket{0}$ also works) which achieves the optimal FI
\begin{equation}
\begin{split}
\max_{\{E(n)\}}\mathcal F_{2\ell+1\,2\ell+1} 
&= 4(2\ell+1)^2\epsilon q_{\ell+1}^2\frac{(M_{2\ell+1})^{4\ell}}{(M_{2\ell})^{4\ell}}.
\end{split}
\end{equation}
Here we use the property that $b^{(\ell)}(x)$ is orthogonal to $\partial^{\ell+1}_{\bar X}\psi_{\rm PSF}(x-\bar X)$ (based on \eqref{eq:prop-1D}). Moreover, according to \eqref{eq:CRbound},
\begin{equation}
\Sigma_{2\ell+1\,2\ell+1} \geq (\mathcal{F}^{-1})_{2\ell+1\,2\ell+1} \geq (\mathcal F_{2\ell+1\,2\ell+1})^{-1}.
\end{equation}
The measurement basis $\ket{\Phi_\pm}$ also leads to $\mathcal F_{2\ell+1\,2\ell} = \mathcal F_{2\ell\,2\ell+1} = O(s^{2\ell})$ which means $\mathcal F$ is effectively diagonal and the second equality in the above inequality holds, because up to the lowest order of $s$ we have
\begin{gather}
Q_{2\ell}(n_+;\{M_k,k\leq 2\ell\}) = Q_{2\ell}(n_-;\{M_k,k\leq 2\ell\}),\\
Q_{2\ell+1}(n_+;\{M_k,k\leq 2\ell+1\}) = - Q_{2\ell+1}(n_-;\{M_k,k\leq 2\ell+1\}),\\
\mathcal F_{2\ell+1\,2\ell} \approx \sum_{\substack{E(n) = \ket{\Phi_+}\bra{\Phi_+}\\E(n) = \ket{\Phi_-}\bra{\Phi_-}}} \frac{1}{Q_{2\ell}(n;\{M_k,k\leq 2\ell\})}\bigg(\frac{\partial Q_{2\ell+1}(n;\{M_k,k\leq 2\ell+1\})}{\partial M_{2\ell+1}}\bigg)\bigg(\frac{\partial Q_{2\ell}(n;\{M_k,k\leq 2\ell\})}{\partial M_{2\ell}}\bigg) = 0.
\end{gather}

\section{Measurement basis and corresponding FIs for weak incoherent sources in 2D imaging}
\label{app:2D}

According to \eqref{eq:q2-even-2D}, by choosing measurement basis 
\begin{equation}
B^{\rm w}_0 = \{b^{(L\,K-L)\dagger}_{{\bar X} {\bar Y}}\ket{0},\;\forall K \geq 0,\; 0 \leq L\leq K\}
\end{equation} 
where $b^{(k\ell)\dagger}_{{\bar X} {\bar Y}} = \int dxdy b^{(k)}_1(x-{\bar X})b^{(\ell)}_2(y-  {\bar Y})a^\dagger_{xy}$, one can achieve the optimal scaling of $s$ (but not necessarily the optimal coefficients) for FIs wrt $M_{2L\,2K-2L}$ for all $K$ and $L\leq K$:
\begin{equation}
\label{eq:2D-even}
\mathcal F_{2L\,2K-2L, 2L\,2K-2L}|_{B^{\rm w}_0} \approx \epsilon q_{L,K-L}^2 (2K)^2 (M_{2L\,2K-2L})^{2K-2} = O(s^{2K -2}).
\end{equation}
By choosing measurement basis 
\begin{equation}
B^{\rm w}_1 = \{\frac{1}{\sqrt{2}}(b^{(L\,K-L)\dagger}_{{\bar X} {\bar Y}}\pm b^{(L+1\,K-L-1)\dagger}_{{\bar X} {\bar Y}}) \ket{0},\;\forall K\geq 0, \;0 \leq L\leq K-1 \text{ is even}\}
\end{equation} (or $B^{\rm w}_2 = \{\frac{1}{\sqrt{2}}(b^{(L\,K-L)\dagger}_{{\bar X} {\bar Y}}\pm b^{(L+1\,K-L-1)\dagger}_{{\bar X} {\bar Y}}) \ket{0},\;\forall K\geq 0,\; 0 \leq L\leq K-1 \text{ is odd}\}$), one can achieve the optimal scaling of $s$ for FIs wrt $M_{2L+1\,2K-(2L+1)}$ for all $K$ is even (or odd) and $L < K$:
\begin{multline}
\mathcal{F}_{2L+1\,2K-(2L+1),2L+1\,2K-(2L+1)}|_{B^{\rm w}_{1,2}} \approx \\ \frac{4\epsilon (q_{L,K-L}^2 (M_{2L,2K-2L})^{2K}+q_{L+1,K-L-1}^2(M_{2L+2,2K-2L-2})^{2K}) q_{L,K-L}^2 q_{L+1,K-L-1}^2(2K)^2(M_{2L+1\,2K-2L-1})^{4K-2}}{(q_{L,K-L}^2 (M_{2L,2K-2L})^{2K}+q_{L+1,K-L-1}^2 (M_{2L+2,2K-2L-2})^{2K})^2-4q_{L,K-L}^2q_{L+1,K-L-1}^2 (M_{2L+1,2K-2L-1})^{4K}} \\= O(s^{2K-2}).
\end{multline}
Meanwhile, $(q_{L,K-L}^2 (M_{2L,2K-2L})^{2K} + q_{L+1,K-L+1}^2 (M_{2L+2,2K-2L-2})^{2K})^{\frac{1}{2K}}$ as a parameter can be estimated simultaneously with precision $O(s^{-2K})$ and independently of $M_{2L+1\,2K-(2L+1)}$. Here we have used the property that $b^{(k \ell)}(x)$ is orthogonal to $\partial_{{\bar X}}^{k'}\partial_{ {\bar Y}}^{\ell'} \psi_{\mathrm{PSF}}(x-{\bar X},y- {\bar Y})$ as long as $k$ and $k'$ (or $\ell$ and $\ell'$) do not have the same parity (i.e. are not both even or odd). 
To conclude, $B^{\rm w}_{1,2}$ cover the estimation of moments whose orders on $x$- and $y$-axis are both even or both odd. The optimal FI scaling is $O(s^{2K-2})$ in this case, where $2K$ is the sum of orders on $x$- and $y$-axis.

For moments who have different parities on $x$- and $y$-axis, we can use basis 
\begin{align}
B^{\rm w}_3 &= \{\frac{1}{\sqrt{2}} (b_{{\bar X}  {\bar Y}}^{(L\,K-L)^\dagger} \pm b_{{\bar X}  {\bar Y}}^{(L+1\,K-L)^\dagger})\ket{0},\;\forall K\geq0,\; 0 \leq L\leq K \text{ is even}\}; \\
B^{\rm w}_4 &= \{\frac{1}{\sqrt{2}} (b_{{\bar X}  {\bar Y}}^{(L\,K-L)^\dagger} \pm b_{{\bar X}  {\bar Y}}^{(L+1\,K-L)^\dagger})\ket{0},\;\forall K\geq0,\; 0 \leq L\leq K \text{ is odd} \}; \\
B^{\rm w}_5 &= \{\frac{1}{\sqrt{2}} (b_{{\bar X}  {\bar Y}}^{(K-L\,L)^\dagger} \pm b_{{\bar X}  {\bar Y}}^{(K-L\,L+1)^\dagger})\ket{0},\;\forall K\geq0,\; 0 \leq L\leq K \text{ is even}\}; \\
B^{\rm w}_6 &= \{\frac{1}{\sqrt{2}} (b_{{\bar X}  {\bar Y}}^{(K-L\,L)^\dagger} \pm b_{{\bar X}  {\bar Y}}^{(K-L\,L+1)^\dagger})\ket{0},\;\forall K\geq0,\; 0 \leq L\leq K \text{ is odd}\}. 
\end{align}
Based on \eqref{eq:q2-odd-2D}, we can calculate the following FIs (up to the lowest order of $s$):
\begin{align}
\mathcal{F}_{2L+1\,2K-2L,2L+1\,2K-2L}|_{B^{\rm w}_{3,4}} &\approx 4\epsilon q_{L+1,K-L}^2(2K+1)^2\frac{(M_{2L+1\,2K-2L})^{4K}}{(M_{2L\,2K-2L})^{2K}} = O(s^{2K});\\
\mathcal{F}_{2K-2L\,2L+1,2K-2L\,2L+1}|_{B^{\rm w}_{5,6}} &\approx 4\epsilon q_{K-L,L+1}^2(2K+1)^2\frac{(M_{2K-2L\,2L+1})^{4K}}{(M_{2K-2L\,2L})^{2K}} = O(s^{2K}).
\end{align}
and $M_{2L\,2K-2L}$ ($M_{2K-2L\,2L}$) can be estimated simultaneously and independently with $M_{2L+1\,2K-2L}$ ($M_{2K-2L\,2L+1}$):
\begin{align}
\mathcal{F}_{2L\,2K-2L,2L\,2K-2L}|_{B^{\rm w}_{3,4}} &\approx \epsilon q_{L,K-L}^2(2K)^2 (M_{2L\,2K-2L})^{2K-2} = O(s^{2K-2});\\
\mathcal{F}_{2K-2L\,2L,2K-2L\,2L}|_{B^{\rm w}_{5,6}} &\approx \epsilon q_{K-L,L}^2(2K)^2 (M_{2K-2L\,2L})^{2K-2} = O(s^{2K-2}).
\end{align}
which are exactly their optimal values as in \eqref{eq:2D-even}.

\section{Estimation of higher order moments with arbitrary source strengths}
\label{app:generalization}

Here we only consider 1D imaging, the discussion can be easily generalized to 2D imaging. 
As already shown in \autoref{sec:second}. Only $0$-null measurement $n\in N_0 =\{n|Q_0(n) =\braket{0|(\psi_{{\bar X}})^k E(n)(\psi_{{\bar X}}^{\dagger})^k|0}=0,\forall k\}$ contributes to the FI wrt $M_2$. Using the method of induction, we define $\ell$-null measurement
\begin{equation}
N_{\ell} = \{n\;|\bra{\Phi}E(n)\ket{\Phi}=0,\forall \ket{\Phi} \in B^{(k)},k\leq \ell\},
\end{equation} 
where $B^{(\ell)} = \{(\prod_k \psi_{{\bar X}}^{(\ell_k)\dagger})\ket{0}, \forall \{\ell_k\geq 0,k\in\mathbb{N}\}\text{, s.t.}\sum_k {\ell_k} = \ell\}$. When $n \in N_{\ell-1}$, $M_{2\ell}$ first apprears in $Q_{2\ell}(n,\{M_k,k\leq 2\ell\})$. Up to the lowest order of $s$,
\begin{equation}
\mathcal F_{2\ell\,2\ell} = \sum_{n}\frac{1}{P(n;\{x_{j},\Gamma_{j}\})}\bigg(\frac{\partial P(n;\{x_{j},\Gamma_{j}\})}{\partial M_{2\ell}}\bigg)^2 \approx \sum_{n \in N_{\ell-1}}\frac{1}{Q_{2\ell}(n,\{M_k,k\leq 2\ell\})}\bigg(\frac{\partial Q_{2\ell}(n,\{M_k,k\leq 2\ell\})}{\partial M_{2\ell}}\bigg)^2,
\end{equation}
where $Q_{2\ell}(n,\{M_k,k\leq 2\ell\})$ is the $O(s^{2\ell})$ order term of 
\begin{equation}
P(n;\{x_{j},\Gamma_{j}\})=\mathbb{E}\Big[\frac{\braket{0|e^{\alpha^{\dagger}\psi}E(n)e^{\psi^{\dagger}\alpha}|0}}{\braket{0|e^{\alpha^{\dagger}\psi}e^{\psi^{\dagger}\alpha}|0}}\Big] = \mathbb{E}[e^{-\int dx |\sum_j\alpha_j\psi_{\mathrm{PSF}}(x-x_j)|^2}\sum_{k=0}^{\infty} \frac{1}{k!^2} \braket{0|(\alpha^\dagger \psi)^k E(n) (\psi^\dagger \alpha)^k|0}].
\end{equation}
When $n\in N_{\ell-1}$, $Q_{2\ell}(n;\{M_k,k\leq 2\ell\})$ has the following form:
\begin{equation}
\label{eq:q2-even-strong}
Q_{2\ell}(n;\{M_k,k\leq 2\ell\}) = \frac{1}{\ell!^2}\sum_{k=0}^{\infty}\frac{\epsilon^{k+1}}{k!(1+\epsilon)^{k+1}}  \braket{0|(\psi_{{\bar X}})^{k}\psi_{{\bar X}}^{(\ell)}E(n)\psi_{{\bar X}}^{(\ell)\dagger}(\psi_{{\bar X}}^{\dagger})^{k}|0}(M_{2\ell})^{2\ell} + Q^{R}_{2\ell}(n;\{M_k,k\leq {2\ell-1}\}),
\end{equation}
where the remainder term $Q^{R}_{2\ell}(n;\{M_k,k\leq 2\ell-1\})$ contains only moments with orders lower than $2\ell$. We note that $Q_{2\ell}(n,\{M_k,k\leq 2\ell\})$ contains only terms like (summing over $k\geq\max\{K^+_0,K^-_0\}$)
\begin{multline}
\frac{1}{(\ell_1^+!\cdots\ell^+_{m_+}!)(\ell_1^-!\cdots\ell^-_{m_-}!)}
 \frac{1}{\big(K^{+}_1!\cdots K^{+}_{m_+}!(k-K^{+}_0)!\big)\big(K^{-}_1!\cdots K^{-}_{m_-}!(k-K^{-}_0)!\big)}\\ \mathbb{E}[e^{-|A^{(0)}|^2}  A^{(0)*k-K^{+}_0} A^{(\ell^{+}_1)*K^{+}_1} \cdots A^{(\ell^{+}_{m_+})*K^{+}_{m_+}}A^{(0)k-K^{-}_0} A^{(\ell^{-}_1)K^{-}_1} \cdots A^{(\ell^{-}_{m_-})K^{-}_{m_-}}] \\\braket{0|(\psi_{{\bar X}})^{k-K^{+}_0}(\psi_{{\bar X}}^{(\ell^{+}_1)})^{K^{+}_1}\cdots (\psi_{{\bar X}}^{(\ell^{+}_{m_+})})^{K^{+}_{m_+}} E(n) (\psi_{{\bar X}}^{(\ell^{-}_1)\dagger })^{K^{-}_1}\cdots (\psi_{{\bar X}}^{(\ell^{-}_{m-})\dagger })^{K^{-}_{m_-}}(\psi_{{\bar X}}^{\dagger })^{k-K^{-}_0}|0},
\end{multline}
where $K^\pm_{m'}, \ell^\pm_{m'}\in \mathbb{N}^+$, $m'=1,\ldots,m_\pm$, $K^\pm_0 = \sum_{m'=1}^{m_\pm} K^\pm_{m'}$ and $\ell = \sum_{m'=1}^{m_\pm} \ell^\pm_m $. From Wick's theorem and \eqref{eq:gaussian-second-moment}, it is clear that 
the only term dependent on $M_{2\ell}$ corresponds to $K^+_0 = K^-_0= 1$, $m_\pm=1$ and $\ell_{m_\pm}=\ell$. 
When $m_\pm=1$ and $\ell_{m_\pm}=\ell$, we have 
\begin{equation}
\mathbb{E}[ e^{-|A^{(0)}|^2} |A^{(0)}|^{2(k-1)}|A^{(\ell)}|^2] = \frac{ (k-1)! \epsilon^k (M_{2\ell})^{2\ell}}{(1+ \epsilon)^{k}} +  \frac{(k-1)!(k-1-\epsilon)\epsilon^k M_{\ell}^{2\ell}}{(1+\epsilon)^{k+1}},
\end{equation}
proving \eqref{eq:q2-even-strong}. Therefore, by choosing the modified measurement
\begin{equation}
\label{eq:basis-0}
B_0 = \Big\{ \sum_{k=0}^\infty \frac{1}{k!}(\psi^\dagger_{{\bar X}})^kb^{(\ell)\dagger}_{{\bar X}}\ket{0}\bra{0}b^{(\ell)}_{{\bar X}}(\psi_{{\bar X}})^k ,\;\forall \ell\geq 0\Big\},
\end{equation}
$Q^{R}_{2\ell}(n;\{M_k,k\leq 2\ell-1\}) = 0$ and the same FI (\eqref{eq:1D-even}) wrt $M_{2\ell}$ is recovered using the modified measurement $B_0$. We can also modify other basis analogously by allowing multi-photon detection of $\psi^\dagger_{{\bar X}}$ and it will provide the same expression of FIs as in the weak source scenario. Note that here each component of $B_0$ is not a POVM but a PVM because we don't need to distinguish the number of $\psi^\dagger_{{\bar X}}$ photon we detect. However if we choose to distinguish them, that is, using
\begin{equation}
{B}'_0 = \Big\{ \frac{1}{k!}(\psi^\dagger_{{\bar X}})^kb^{(\ell)\dagger}_{{\bar X}}\ket{0}\bra{0}b^{(\ell)}_{{\bar X}}(\psi_{{\bar X}})^k ,\;\forall k,\ell\geq 0\Big\},
\end{equation}
the FI will be no smaller (easily proven using Cauchy-Schwarz inequality) and the FIM is still effectively diagonal.

However, even ${B}'_0$ is not optimal when estimating $M_{2\ell}$. 
Physically, the reason is that the information of high order moments can be obtained by detecting several low order derivative operators simultaneously, which is neglectable when the source is weak. For $\ell=1$, the only lower order moment is $M_{1} = 0$, therefore strong source strength does not make a difference when calculating the FI, as shown in \autoref{sec:second}. 

We provide a simple example showing ${B}'_0$ is not an optimal measurement basis by replacing it with a better basis. Consider $\ell=4$ (and we want to estimate the value of $M_{2\ell} = M_{8}$). Suppose $s>0$. 
For simplicity, we only consider the replacement in $2$-photon subspace, i.e. we don't change any $k+1$-photon basis in ${B}'_0$ with $k\neq 1$ and their contributions to $\mathcal F_{88}$ will remain the same. For $2$-photon subspace, we consider the possiblity of choosing another basis in $B_{4,2^2} = \mathrm{span}\{\psi_{{\bar X}}^{\dagger} b^{(4)\dagger}  \ket 0, \frac{1}{\sqrt{2}} b^{(2)\dagger} b^{(2)\dagger}  \ket 0\} \equiv \mathrm{span}\{\gv{b_4},\gv{b_{2^2}}\}$. After some calculations, we have $Q_{8}(n;\{M_{k},k\leq 8 \})$ in this 2-dimensional subspace
\begin{equation}
Q_{8}(n;\{M_{k},k\leq 8 \}) = \trace\bigg(E(n)\begin{pmatrix}A_{44}&A_{42^2}\\A_{2^24}&A_{2^22^2}\end{pmatrix}\bigg),
\end{equation}
where
\begin{gather}
A_{44} =  q_4^2\mathbb{E}[e^{- \abs{A^{(0)}}^2} |A^{(4)}|^2 |A^{(0)}|^2] 
= q_4^2\frac{\epsilon^2}{(1+\epsilon)^2} \Big( (M_{8}^8 - M_{4}^8) + \frac{2}{1+\epsilon} M_4^8 \Big) > 0,\\
A_{42^2} = A_{2^24} =  q_4\Big(\frac{q_2^2}{4}\Big) \mathbb{E}[e^{- \abs{A^{(0)}}^2} A^{(4)}A^{(0)}(A^{(2)*})^2])
= q_4\Big(\frac{q_2^2}{4}\Big) \frac{2\epsilon^2}{(1+\epsilon)^2} \Big( (M^6_{6}M^2_{2} - M^4_{4}M^4_{2}) + \frac{1}{1+\epsilon} M^4_{4}M^4_{2} \Big) > 0, \\
A_{2^22^2} =  \Big(\frac{q_2^2}{4}\Big)^2 \mathbb{E}[e^{- \abs{A^{(0)}}^2} |A^{(2)}|^4] 
= \Big(\frac{q_2^2}{4}\Big)^2 \Big( \frac{2\epsilon^2}{1+\epsilon}M_4^8 - \frac{4\epsilon^3}{(1+\epsilon)^2}M_4^4M^4_{2} + \frac{2\epsilon^4}{(1+\epsilon)^3} M_2^8\big) > 0.
\end{gather}
We can easily find a non-trivial image such that $A_{44}A_{2^22^2} - A_{42^2}^2 >0$, then we maximize $\mathcal F_{88}$ in this 2-dimensional subspace by doing QFI calculation, which gives
\begin{multline}
\max_{E(n) \text{ in } B_{4,2^2}}\frac{1}{Q_{8}(n;\{M_{k},k\leq 8 \})}\bigg(\frac{\partial Q_{8}(n;\{M_{k},k\leq 8 \})}{\partial M_8}\bigg)^2 = \bigg(q_4^28M_8^7\frac{\epsilon^2}{(1+\epsilon)^2}\bigg)^2\frac{}{}\frac{A_{44}A_{2^22^2} - A_{42^2}^2 + A_{2^22^2}^2}{(A_{44}+A_{2^22^2})(A_{44}A_{2^22^2} - A_{42^2}^2)}\\
 > \frac{1}{Q_{8}(n;\{M_{k},k\leq 8 \})}\bigg(\frac{\partial Q_{8}(n;\{M_{k},k\leq 8 \})}{\partial M_8}\bigg)^2\bigg|_{E(n) = \gv{b_4} \gv{b_4}^\dagger} = \bigg(q_4^28M_8^7\frac{\epsilon^2}{(1+\epsilon)^2}\bigg)^2\frac{1}{A_{44}}.
\end{multline}
Now we've proven $\gv{b_4} \gv{b_4}^\dagger$ does not generate the maximum FI wrt $M_8$ and ${B}'_0$ is not optimal. Meanwhile, we also note that the FIM is effectively diagonal in the subdiffraction limit, thus $\mathcal F_{88}$ fully characterizes the measurement precision of $M_8$. 
In general, any non-zero off-diagonal term ($A_{42^2}$ in this case) in the same photon number subspace would lead to the same result. It means the precision of high order moments estimation could be enhanced by utilizing the detection of several low order derivative operators simultaneously. 

For odd moments $M_{2\ell+1}$ ($\ell \geq 1$), suppose $n \in N_{\ell-1}\backslash N_{\ell+1}$, we have
\begin{multline}
Q_{2\ell+1}(n;\{M_k,k\leq 2\ell+1\}) = \\ \frac{1}{\ell!(\ell+1)!}\sum_{k=0}^{\infty}\frac{\epsilon^{k+1}}{k!(1+\epsilon)^{k+1}}  \mathrm{Re}[\braket{0|(\psi_{{\bar X}})^{k}\psi_{{\bar X}}^{(\ell+1)}E(n)\psi_{{\bar X}}^{(\ell)\dagger}(\psi_{{\bar X}}^{\dagger})^{k}|0}] (M_{2\ell+1})^{2\ell+1} + Q^{R}_{2\ell+1}(n;\{M_k,k\leq {2\ell}\}).
\end{multline}
The modified measurement 
\begin{equation}
\label{eq:basis-1}
B_{1(2)} = \Big\{ \sum_{k=0}^\infty \frac{1}{k!}(\psi^\dagger_{{\bar X}})^k\frac{b^{(\ell)\dagger}_{{\bar X}}\pm b^{(\ell+1)\dagger}_{\bar X}}{\sqrt{2}}\ket{0}\bra{0}\frac{b^{(\ell)}_{{\bar X}}\pm b^{(\ell+1)}_{\bar X}}{\sqrt{2}}(\psi_{{\bar X}})^k ,\;\forall \ell \text{ is odd (or even)}\Big\},
\end{equation}
also leads to the same FI \eqref{eq:1D-odd}, as for the even moments.



\section{Pre-estimation of the centroid}
\label{app:centroid}

The procedure to estimate the centroid can be divided into two step: (1) Find a reference point $X_R$ such that $\abs{\bar{X}-X_R} \lesssim s$; (2) Precisely locate $\bar X$ within the subdiffraction limit. The resource required for step (1) is neglectable (it's a coarse estimation) and we only consider the resource required for step (2). Normally, to fully resolve an image, we need to achieve a degree of precision where $\delta M_k \ll s$ ($k \geq 2$) and here we analyze the resource required to achieve $\delta \bar X \ll s$ so that it won't induce a significant error in the estimation of higher other moments. 

We first consider 1D weak source scenario. After step (1), we are already in the subdiffraction regime and we can expand $P(n;\{x_j,\Gamma_j\})$ around $X_R$ up to $O(s^2)$, which gives
\begin{multline}
P(n;\{x_j,\Gamma_j\}) \approx Q_0(n)+Q_1(n)+Q_2(n) = \epsilon\bra{0}\psi_{X_R}E(n)\psi_{X_R}^\dagger\ket{0} + 2\epsilon\mathrm{Re}[\bra{0}\psi_{X_R}^{(1)}E(n)\psi_{X_R}^\dagger\ket{0}]\tilde M_1 \\ + \epsilon(\bra{0}\psi^{(1)}_{X_R} E(n) \psi^{(1)\dagger}_{X_R}\ket{0}+2\mathrm{Re}[\bra{0}\psi^{(2)}_{X_R} E(n) \psi^{\dagger}_{X_R}\ket{0}])\tilde M_2^2 + O(\epsilon^2).
\end{multline}
Here $\tilde M_1$ and $\tilde M_2$ is redefined using $X_R$ as the centroid. According to \appref{app:odd-moments}, the optimal measurement in terms of estimating $\bar M_1 = \bar X - X_R$ can be an arbitrary projection onto two orthonormal basis in the real span of $\{\psi_{X_R}^\dagger\ket{0},\psi_{X_R}^{(1)\dagger}\ket{0}\}$ as long as $Q_0(n)\gg Q_1(n) \gg Q_2(n)$ is satisfied. For example,
\begin{equation}
E(n_\pm) = \Big(\frac{\psi_{X_R}^\dagger \pm \frac{1}{\Delta k}\psi_{X_R}^{(1)\dagger} }{\sqrt{2}}\Big)\ket{0}\bra{0}\Big(\frac{\psi_{X_R} \pm \frac{1}{\Delta k}\psi_{X_R}^{(1)} }{\sqrt{2}}\Big)
\end{equation}
is optimal. The corresponding FI is 
\begin{equation}
\label{eq:fi1-weak}
\mathcal F_{11} = 4\epsilon \Delta k^2.
\end{equation}
which is the same as \eqref{eq:fi2-weak}. Therefore, if we want to estimate both the second moment $M_2$ and the centroid $\bar X$, a straightforward method is to first use half of the whole resource to locate $\bar X$ such that $\delta \bar X \ll s$ and then use the rest half to estimate $M_2$ as described in \autoref{sec:second}. The effective FIM would be half of the optimal ones \eqref{eq:fi1-weak} and \eqref{eq:fi2-weak},
\begin{equation}
\label{eq:fi12-weak}
\mathcal F(\tilde{M}_1, M_2) = \begin{pmatrix} 2\epsilon \Delta k^2 & 0 \\ 0 & 2\epsilon \Delta k^2\end{pmatrix},
\end{equation}
which is only half of the QFIM \cite{tsang2016quantum}
\begin{equation}
\label{eq:qfi12-weak}
\mathcal J(\tilde{M}_1, M_2) = \begin{pmatrix} 4\epsilon \Delta k^2 & 0 \\ 0 & 4\epsilon \Delta k^2\end{pmatrix}.
\end{equation}
When we want to estimate even higher order moments, the resource required to locate $\bar{X}$ is neglectable. 

Now we show \eqref{eq:fi12-weak} is the optimal precision we can get (in the subdiffraction limit) and the QFIM \eqref{eq:qfi12-weak} is not achievable. For any POVM $\{E(n)\}$, the only case when $P(n,\{x_j,\Gamma_j\})$ does not lead to a zero-FI wrt $M_2$ is when there is an $E(n)$ such that
\begin{equation}
P(n,\{x_j,\Gamma_j\}) \approx  A_0(n) + A_1(n) \tilde{M_1} + A_2(n) (M_2^2 + \tilde M_1^2) = O(s^2)
\end{equation}
where 
\begin{align}
A_0(n) &= \epsilon\bra{0}\psi_{X_R}E(n)\psi_{X_R}^\dagger\ket{0} = O(s^2), \\
A_1(n) &= 2\epsilon\mathrm{Re}[\bra{0}\psi_{X_R}^{(1)}E(n)\psi_{X_R}^\dagger\ket{0}] = O(s), \\
A_2(n) &= \epsilon(\bra{0}\psi^{(1)}_{X_R} E(n) \psi^{(1)\dagger}_{X_R}\ket{0}+2\mathrm{Re}[\bra{0}\psi^{(2)}_{X_R} E(n) \psi^{\dagger}_{X_R}\ket{0}]) =O(1),
\end{align}
and we use the relation $\tilde M_2^2  = M_2^2 + M_1^2$. Note that $A_2(n) \approx \epsilon\bra{0}\psi^{(1)}_{X_R} E(n) \psi^{(1)\dagger}_{X_R}\ket{0}$, because $\mathrm{Re}[\bra{0}\psi^{(2)}_{X_R} E(n) \psi^{\dagger}_{X_R}\ket{0}] = O(s)$ can be negelected. Since
\begin{align}
\frac{1}{P(n,\{x_j,\Gamma_j\})}\bigg(\frac{\partial P(n,\{x_j,\Gamma_j\})}{\partial \tilde M_1}\bigg)^2 &= \frac{(A_1(n)+2A_2(n)\tilde M_1)^2}{A_0(n) + A_1(n) \tilde{M_1} + A_2(n) (M_2^2 + \tilde M_1^2)},\\
\frac{1}{P(n,\{x_j,\Gamma_j\})}\bigg(\frac{\partial P(n,\{x_j,\Gamma_j\})}{\partial M_2}\bigg)^2 &= \frac{4A_2(n)^2M_2^2}{A_0(n) + A_1(n) \tilde{M_1} + A_2(n) (M_2^2 + \tilde M_1^2)}
\end{align}
and $A_1(n)^2 \leq 4 A_2(n) A_0(n)$, we have
\begin{equation}
\label{eq:appEproof}
\frac{1}{P(n,\{x_j,\Gamma_j\})}\bigg(\frac{\partial P(n,\{x_j,\Gamma_j\})}{\partial M_2}\bigg)^2 +
\frac{1}{P(n,\{x_j,\Gamma_j\})}\bigg(\frac{\partial P(n,\{x_j,\Gamma_j\})}{\partial \tilde M_1}\bigg)^2 \lesssim 4A_2(n) \approx 4\epsilon\bra{0}\psi^{(1)}_{X_R} E(n) \psi^{(1)\dagger}_{X_R}\ket{0}.
\end{equation}
When $P(n,\{x_j,\Gamma_j\})$ is dominated by $Q_0(n)$, we also have
\begin{equation}
\frac{1}{P(n,\{x_j,\Gamma_j\})}\bigg(\frac{\partial P(n,\{x_j,\Gamma_j\})}{\partial M_2}\bigg)^2 +
\frac{1}{P(n,\{x_j,\Gamma_j\})}\bigg(\frac{\partial P(n,\{x_j,\Gamma_j\})}{\partial \tilde M_1}\bigg)^2 \approx \frac{A_1(n)^2}{A_0(n)} \leq 4\epsilon\bra{0}\psi^{(1)}_{X_R} E(n) \psi^{(1)\dagger}_{X_R}\ket{0}. 
\end{equation}
Therefore, any achievable FIM must satisfies
\begin{equation}
\mathcal F_{11} + \mathcal F_{22} \leq 4\epsilon\sum_n \bra{0}\psi^{(1)}_{X_R} E(n) \psi^{(1)\dagger}_{X_R}\ket{0} = 4\epsilon \Delta k^2,
\end{equation}
and 
\begin{equation}
(\delta \tilde{M}_1)^2 + (\delta M_2)^2 = \trace(\Sigma) \geq \trace(\mathcal F^{-1}) \geq \sum_{i=1}^2 \mathcal F_{ii}^{-1} \geq \frac{4}{\trace(\mathcal F)} \geq \frac{1}{\epsilon\Delta k^2}.
\end{equation}
Clearly the last three equalities are simultaneously satisfied when FIM is \eqref{eq:fi12-weak}, implying the optimality of our measurement scheme.

The situation becomes a bit more complicated for arbitrary source strengths. First, we expand $P(n;\{x_j,\Gamma_j\})$ around $X_R$ up to $O(s)$
\begin{multline}
P(n;\{x_j,\Gamma_j\}) \approx Q_0(n)+Q_1(n) =  \sum_{k=0}^{\infty} \frac{\epsilon^k}{k!(1+\epsilon)^{k+1}} \braket{0|\psi_{{X_R}}^k E(n) (\psi_{{X_R}}^\dagger)^k|0} \\ + \sum_{k=0}^\infty \frac{2 \epsilon^{k+1}}{k!(1+\epsilon)^{k+2}} \mathrm{Re}[\braket{0|\psi_{{X_R}}^{k}\psi_{{X_R}}^{(1)} E(n) (\psi_{{X_R}}^\dagger)^{k+1}|0}] \tilde M_1.
\end{multline}
Since the quantum state is photon number diagonal, the optimal measurement estimating $\tilde M_1$ must also be photon number diagonal \cite{braunstein1994statistical}, that is, $\{E(n)\}$ should contains $\{E(n_k), k\geq 1\}$ where $E(n_k) = \Pi_{\v k}E(n_k)\Pi_{\v k}$ and $\Pi_{\v k}$ is projection onto $\v k$-photon subspace. In this case, we shall write 
\begin{equation}
\label{eq:fi1}
\mathcal F_{11} = \sum_{k=0}^\infty \sum_{\{E(n_k)\}} \frac{\big(\frac{2 \epsilon^{k+1}}{k!(1+\epsilon)^{k+2}} \mathrm{Re}[\braket{0|\psi_{{X_R}}^{k}\psi_{{X_R}}^{(1)} E(n_{k+1}) (\psi_{{X_R}}^\dagger)^{k+1}|0}]\big)^2 }{\big(\frac{\epsilon^{k+1}}{(k+1)!(1+\epsilon)^{k+2}} \braket{0|\psi_{{X_R}}^{k+1} E(n_{k+1}) (\psi_{{X_R}}^\dagger)^{k+1}|0}\big)} \leq 4\epsilon \Delta k^2,
\end{equation}
where the equality holds when $\{E(n_k)\}$ is an arbitrary projection onto two orthonormal basis in the real span of $\{(\psi_{X_R}^\dagger)^{k-1}\psi_{X_R}^{(1)\dagger}\ket{0},(\psi_{X_R}^\dagger)^{k}\ket{0}\}$ as long as $Q_0(n)\gg Q_1(n) \gg Q_2(n)$ is satisfied. 
For example,
\begin{equation}
E(n_{k,\pm}) = \frac{1}{2}\Big(\frac{1}{\sqrt{k!}}(\psi_{X_R}^\dagger)^{k} \pm \frac{1}{\Delta k\sqrt{(k-1)!}}\psi_{X_R}^{(1)\dagger}(\psi_{X_R}^\dagger)^{k-1} \Big)\ket{0}\bra{0}\Big(\frac{1}{\sqrt{k!}}\psi_{X_R}^{k} \pm \frac{1}{\Delta k\sqrt{(k-1)!}}\psi_{X_R}^{(1)}\psi_{X_R}^{k-1} \Big)
\end{equation}
is optimal. Therefore, if we want to estimate both the second moment $M_2$ and the centroid $\bar X = \tilde M_1 + X_R$, a straightforward method is to first use half of the whole resource to locate $\bar X$ such that $\delta \bar X \leq s$ and then use the rest half to estimate $M_2$ as described in \autoref{sec:second}. Note that to achieve the optimal precision wrt $M_1$, one has to count the number of detected photons by projecting the quantum state onto 
\begin{equation}
\tilde B = \Big\{\frac{1}{2}\Big(\frac{1}{\sqrt{k!}}(\psi_{X_R}^\dagger)^{k} \pm \frac{1}{\Delta k\sqrt{(k-1)!}}\psi_{X_R}^{(1)\dagger}(\psi_{X_R}^\dagger)^{k-1} \Big)\ket{0}\bra{0}\Big(\frac{1}{\sqrt{k!}}\psi_{X_R}^{k} \pm \frac{1}{\Delta k\sqrt{(k-1)!}}\psi_{X_R}^{(1)}\psi_{X_R}^{k-1} \Big), k \geq 1\Big\}
\end{equation}
unlike using \eqref{eq:strong-measurement} to estimate $M_2$ where we don't need to count the number of photons. Similar to the weak soure scenario, this measurement scheme provides an effective FIM which is half of the optimal ones \eqref{eq:fi1} and \eqref{eq:fi2},
\begin{equation}
\label{eq:fi12}
\mathcal F(\tilde{M}_1, M_2) = \begin{pmatrix} 2\epsilon \Delta k^2 & 0 \\ 0 & 2\epsilon \Delta k^2\end{pmatrix}. 
\end{equation}
It is only half of the QFIM 
\begin{equation}
\label{eq:qfi12}
\mathcal J(\tilde{M}_1, M_2) = \begin{pmatrix} 4\epsilon \Delta k^2 & 0 \\ 0 & 4\epsilon \Delta k^2\end{pmatrix}.
\end{equation}
The resource required to locate $\bar X$ when we want to estimate even higher order moments is still neglectable as in the weak source scenario. Now we consider the possiblity of further improving \eqref{eq:fi12}, here we show that above scheme is at least 96.4\% efficient. 
According to \appref{app:expansion}, we have, up to $O(s^2)$, 
\begin{equation}
P(n,\{x_j,\Gamma_j\}) = A_0(n) + A_1(n) \tilde M_1 + A_2(n) M_2^2 + A_3(n) \tilde M_1^2.
\end{equation}
For different measurement outcome $n$, there are only two situations:
\begin{itemize}
\item $P(n,\{x_j,\Gamma_j\}) = O(s^2)$, then 
\begin{align}
A_0(n) &= \sum_{k=0}^\infty\frac{\epsilon^{k+1}}{(k+1)!(1+\epsilon)^{k+2}} \braket{0|\psi_{{X_R}}^{k+1} E(n) (\psi_{{X_R}}^\dagger)^{k+1}|0}, \\
A_1(n) &= \sum_{k=0}^\infty\frac{2 \epsilon^{k+1}}{k!(1+\epsilon)^{k+2}} \mathrm{Re}[\braket{0|\psi_{{X_R}}^{k}\psi_{{X_R}}^{(1)} E(n) (\psi_{{X_R}}^\dagger)^{k+1}|0}], \\
A_2(n) &= \sum_{k=0}^\infty\frac{\epsilon^{k+1}}{k!(1+\epsilon)^{k+1}}\braket{0|\psi_{{X_R}}^{k}\psi_{{X_R}}^{(1)} E(n) \psi_{{X_R}}^{(1)\dagger}(\psi_{{X_R}}^\dagger)^{k}|0}, \label{eq:appEproof2}\\
A_3(n) &= \sum_{k=0}^\infty\frac{(k+1)\epsilon^{k+1}}{k!(1+\epsilon)^{k+2}}\braket{0|\psi_{{X_R}}^{k}\psi_{{X_R}}^{(1)} E(n) \psi_{{X_R}}^{(1)\dagger}(\psi_{{X_R}}^\dagger)^{k}|0}. \label{eq:appEproof3}
\end{align}
Other terms can be ignored in the subdiffraction limit.
\item $P(n,\{x_j,\Gamma_j\}) = O(1)$, then
\begin{align}
A_0(n) &= \sum_{k=0}^\infty\frac{\epsilon^{k+1}}{(k+1)!(1+\epsilon)^{k+2}} \braket{0|\psi_{{X_R}}^{k+1} E(n) (\psi_{{X_R}}^\dagger)^{k+1}|0}, \\
A_1(n) &= \sum_{k=0}^\infty\frac{2 \epsilon^{k+1}}{k!(1+\epsilon)^{k+2}} \mathrm{Re}[\braket{0|\psi_{{X_R}}^{k}\psi_{{X_R}}^{(1)} E(n) (\psi_{{X_R}}^\dagger)^{k+1}|0}],
\end{align} 
and $A_2(n)$ and $A_3(n)$ can be ignored in the subdiffraction limit. For simplicity we can assume \eqref{eq:appEproof2} and \eqref{eq:appEproof3} are also true. 
\end{itemize}
One important property derived from this relation is that
\begin{equation}
\label{eq:appEproof4}
\sum_{n} A_2(n) = \sum_{n} A_3(n) = \epsilon\Delta k^2.
\end{equation} 
The entries of the FIM are
\begin{gather}
\mathcal F_{11} = \sum_{n} \frac{(2A_3(n)M_1')^2}{A'_0(n) + A_2(n) M_2^2 + A_3(n) M_1'^2};\\
\mathcal F_{12} = \mathcal F_{21} = \sum_{n} \frac{(2A_3(n)M_1')(2A_2(n)M_2)}{A'_0(n) + A_2(n) M_2^2 + A_3(n) M_1'^2};\\ 
\mathcal F_{22} = \sum_{n} \frac{(2A_2(n)M_2)^2}{A'_0(n) + A_2(n) M_2^2 + A_3(n) M_1'^2}.
\end{gather} 
where $M_1' = \tilde M_1 + A_1(n)/(2A_3(n))$ and $A'_0(n) = A_0(n) - A_1(n)^2/(4A_3(n)) \geq 0$. We define another 2-by-2 matrix $\mathcal F'$ by replacing all $A_0(n)$ above with 0. Clearly, $\trace(\mathcal F^{-1}) \geq \trace(\mathcal F'^{-1})$ because $\mathcal F' \succeq \mathcal F$. 
Using \eqref{eq:appEproof4}, we have
\begin{equation}
\mathcal F'_{11} + \frac{M_2}{M_1'} \mathcal F'_{12} = \mathcal F'_{22} + \frac{M_1'}{M_2} \mathcal F'_{12} = 4\epsilon\Delta k^2,
\end{equation}
and
\begin{equation}
\begin{split}
\mathcal F'_{11} + \mathcal F'_{22} &\leq 4\sum_{n}\max\{A_2(n),A_3(n)\}  \\
&\leq 4 \Delta k^2 \Big(\sum_{k=0}^{\lfloor \epsilon \rfloor} \frac{\epsilon^{k+1}}{(1+\epsilon)^{k+1}} + \sum_{k=\lfloor \epsilon \rfloor+1}^\infty\frac{(k+1)\epsilon^{k+1}}{(1+\epsilon)^{k+2}}\Big) 
\\& = 4 \Delta k^2 \Big(\epsilon + \lfloor \epsilon + 1\rfloor \big(\frac{\epsilon}{1+\epsilon}\big)^{\lfloor \epsilon + 2 \rfloor }\Big) \leq 4(1+\frac{1}{e})\epsilon\Delta k^2.
\end{split}
\end{equation}
Therefore 
\begin{equation}
\begin{split}
(\delta \tilde M_1)^2 + (\delta \tilde M_2)^2 &\geq \trace(\mathcal F^{-1}) \geq \trace(\mathcal F'^{-1}) = \frac{\mathcal F'_{11} + \mathcal F'_{22}}{\mathcal F'_{11}\mathcal F'_{22} - \mathcal F'^2_{12}}\\ &   = \frac{1}{4\epsilon \Delta k^2 \big(1- \frac{4\epsilon \Delta k^2}{\mathcal F'_{11} + \mathcal F'_{22}}\big)} \geq \frac{1+e}{4}\frac{1}{\epsilon \Delta k^2}.
\end{split}
\end{equation}
We conclude that our measurement scheme is at least $\sqrt{\frac{1+e}{4}}\approx 96.4\%$ efficient for arbitrary $\epsilon$ in the sense that if one achieve certain estimation precision $\sqrt{(\delta \tilde M_1)^2 + (\delta \tilde M_2)^2}$ by repeating our measurement $N$ times, the optimal measurement scheme requires at least $96.4\%\cdot N$ times to achieve such precision.

We can easily generalize above measurement scheme to 2D imaging when the PSF is separable. $\psi_{X_R Y_R}^{(10)\dagger}$ and $\psi_{X_R Y_R}^{(01)\dagger}$ are orthogonal. As in 1D imaging, 
\begin{equation} 
\tilde M_{10} = \bar X - X_R, \qquad \tilde M_{01} = \bar Y - Y_R
\end{equation}
are estimated by
\begin{equation} 
\frac{1}{k!}(\psi^\dagger_{{X_R Y_R}})^k\frac{\psi_{X_R Y_R}^\dagger \pm \frac{1}{\Delta k_x} \psi_{X_R Y_R}^{(10)\dagger}}{\sqrt{2}}\ket{0}\bra{0}\frac{\psi_{X_R Y_R} \pm \frac{1}{\Delta k_x} \psi_{X_R Y_R}^{(10)}}{\sqrt{2}}(\psi_{{X_R Y_R}})^k
\end{equation}
and
\begin{equation}
\frac{1}{k!}(\psi^\dagger_{{X_R Y_R}})^k\frac{\psi_{X_R Y_R}^\dagger \pm \frac{1}{\Delta k_y} \psi_{X_R Y_R}^{(01)\dagger}}{\sqrt{2}}\ket{0}\bra{0}\frac{\psi_{X_R Y_R} \pm \frac{1}{\Delta k_y} \psi_{X_R Y_R}^{(01)}}{\sqrt{2}}(\psi_{{X_R Y_R}})^k
\end{equation}
with optimal FIs equal to
\begin{equation} 
\mathcal F_{10\,10} = 4\epsilon \Delta k_x^2,\qquad \mathcal F_{01\,01} = 4\epsilon \Delta k_y^2. 
\end{equation}
We won't discuss simultaneous estimation of the centroid $\tilde M_{10}, \tilde M_{01}$ and the second moments $M_{20},M_{11},M_{02}$ here. 


\end{document}